\documentclass[aps,prl,twocolumn,superscriptaddress,floatfix,citeautoscript,preprintnumbers,raggedbottom]{revtex4-2}

\usepackage{graphicx}
\usepackage{amssymb,amsmath,amsthm,mathrsfs,amsfonts,mathtools,comment}
\usepackage{hyperref}
\hypersetup{
  colorlinks,
  citecolor=blue,
  filecolor=blue,
  linkcolor=blue,
  urlcolor=blue
}
\usepackage{aliascnt}
\usepackage[capitalise]{cleveref}
\usepackage{bm}
\usepackage{physics}
\usepackage{braket}
\usepackage[flushleft]{threeparttable}
\usepackage{booktabs}

\usepackage{makecell}
\usepackage{caption}
\usepackage[labelfont=bf]{subcaption}

\newlength{\fskip}
\newlength{\fgap}
\usepackage{dsfont}
 \usepackage{multirow}
 \newcommand{\dInt}{{\mathrm{d}}}

\renewcommand{\tr}{{\mathrm{Tr}}}

\renewcommand{\norm}[1]{\lVert #1 \rVert}

\newcommand{\kett}[1]{\lvert #1\rangle\!\rangle}
\newcommand{\bbra}[1]{\langle\!\langle #1\rvert}
 \newcommand{\bP}{\boldsymbol{P}}
 \newcommand{\I}{\mathrm{i}}
\newcommand{\mc}{\mathcal}
\newcommand{\wt}{\widetilde}

\newcommand{\bv}{\mathbf}
\newcommand{\bs}{\boldsymbol}

\renewcommand{\Re}{\operatorname{Re}}
\renewcommand{\Im}{\operatorname{Im}}

\newcommand{\tnorm}[1]{{\left\vert\kern-0.25ex\left\vert\kern-0.25ex\left\vert#1\right\vert\kern-0.25ex\right\vert\kern-0.25ex\right\vert}}

\newcommand{\RR}{\mathbb{R}}
\newcommand{\CC}{\mathbb{C}}

\newcommand{\ZZ}{\mathbb{Z}}

\newaliascnt{prop}{thm}
\newtheorem{prop}[prop]{\protect\propositionname}
\aliascntresetthe{prop}
\crefname{prop}{Proposition}{Propositions}
\crefname{section}{Sec.}{Secs.}
\Crefname{section}{Sec.}{Secs.}
\crefname{subsection}{Sec.}{Secs.}
\Crefname{subsection}{Sec.}{Secs.}
\Crefname{prop}{Proposition}{Propositions}

\providecommand{\definitionname}{Definition}
\providecommand{\assumptionname}{Assumption}
\providecommand{\corollaryname}{Corollary}
\providecommand{\lemmaname}{Lemma}
\providecommand{\propositionname}{Proposition}
\providecommand{\remarkname}{Remark}
\providecommand{\theoremname}{Theorem}

\makeatletter
\renewcommand*\env@matrix[1][\arraystretch]{%
        \edef\arraystretch{#1}%
        \hskip -\arraycolsep
        \let\@ifnextchar\new@ifnextchar
        \array{*\c@MaxMatrixCols c}}
\makeatother
\usepackage{bibunits}
\usepackage{etoolbox}

\begin{document}

\title{
Efficient simulations of spectra and transport for multiorbital quantum impurities
}

\begin{abstract}
Quantum impurity calculations become challenging when multiple correlated orbitals are treated explicitly, particularly in nonequilibrium steady states. We introduce Coupled-Cluster Coupled Lindblad (C$^3$L), a deterministic approach that combines the recently developed coupled-Lindblad compact bath representation with coupled-cluster theory to compute steady states and Green's functions directly in imaginary and real frequency. C$^3$L has favorable polynomial computational complexity and avoids the sign problem of quantum Monte Carlo (QMC). In a real-time multiorbital benchmark, it agrees closely with state-of-the-art QMC while reducing computational cost by four to five orders of magnitude, with a further 10--100-fold speedup on a GPU. Beyond the practical reach of established real-time QMC solvers, four-orbital nonequilibrium impurities with general interactions and biased reservoirs become accessible with C$^3$L. We obtain resonance-resolved transport signatures relevant to scanning tunneling microscopy experiments on molecular junctions.
\end{abstract}

 \newcommand{\berkeleymath}{Department of Mathematics, University of California, Berkeley, CA 94720, USA}
 \newcommand{\ccq}{Center for Computational Quantum Physics, Flatiron Institute, New York, NY 10010, USA}
 \newcommand{\ccm}{Center for Computational Mathematics, Flatiron Institute, New York, NY 10010, USA}
 \newcommand{\caltechcms}{Department of Computing and Mathematical Sciences, California Institute of Technology, Pasadena, CA 91125, USA}
 \newcommand{\lbnl}{Applied Mathematics and Computational Research Division, Lawrence Berkeley National Laboratory, Berkeley, CA 94720, USA}

\author{Hao-En Li}
\email{haoen2021@math.berkeley.edu}
\affiliation{\berkeleymath}
\author{Zhen Huang}
\email{zhuang@flatironinstitute.org}
\affiliation{\ccq}
\affiliation{\ccm}
\author{Lin Lin}
\email{lin@caltech.edu}
\affiliation{\berkeleymath}
\affiliation{\caltechcms}
\affiliation{\lbnl}

\date{October 28, 2026}
\maketitle

\emph{Introduction.}---%
Quantum impurity models describe the dynamics of correlated electronic orbitals in quantum matter and charge transport through mesoscopic devices by treating an interacting region coupled to its environment as an open quantum system~\cite{SavrasovHauleOudovenkoEtAl2006,ThossEvers2018,EversKorytarTewariVanRuitenbeek2020,MaurerKonigSchoeller2020}. The same framework also underpins quantum embedding approaches to correlated materials~\cite{GeorgesKotliarKrauthRozenberg1996,AokiTsujiEckstein2014,ErpenbeckBlommelZhangEtAl2024,DeMediciMravljeGeorges2011,EidelsteinGullCohen2020}. Accounting for general multiorbital interactions at practical cost, both in and out of equilibrium, would greatly expand the range of phenomena accessible to impurity solvers.

Treating several correlated orbitals remains challenging for established impurity solvers. Continuous-time quantum Monte Carlo (CT-QMC), including interaction-expansion (CT-INT) and hybridization-expansion (CT-HYB), provides numerically exact results but can encounter severe sign problems in models with off-diagonal hybridization and general interactions~\cite{SethKrivenkoFerrero2016}. Quantics tensor trains accelerate deterministic diagrammatic expansions, but the difficulty of compressing multiorbital integrands has so far restricted this approach to single-orbital models~\cite{KimWerner2025,MatsuuraShinaokaWernerTsuji2025}. Separable imaginary-time diagram methods reach realistic three-band models but are most applicable to strong-coupling regimes~\cite{KayeHuangStrandGolez2024,HuangKayeStrandGolez2025}. Even in equilibrium, a few correlated orbitals with general interactions can already be prohibitively expensive.

Dynamical and transport properties are even more challenging to obtain. Equilibrium calculations typically compute imaginary-time Green's functions, and obtaining real-frequency spectra requires ill-posed analytic continuation. Nonequilibrium properties instead call for real-time or Keldysh-contour approaches~\cite{AokiTsujiEckstein2014}. Many-body perturbation theories (MBPTs) like self-consistent second-Born (2ndB) and $GW$ approximations offer a practical treatment of nonequilibrium Green's functions, but their selected self-energy diagrams can miss qualitative features~\cite{SpataruHybertsenLouieMillis2009,MyohanenStanStefanucciEtAl2009}. Real-time QMC suffers from a dynamical sign problem in long-time simulations~\cite{WernerOkaMillis2009,CohenGullReichmanMillis2015}. Inchworm methods reach longer times but remain extremely costly for general multiorbital systems~\cite{ErpenbeckGullCohen2023,ErpenbeckBlommelZhangEtAl2024}.  The fork tensor-product-state solver avoids sign problems, but its efficiency relies on  orbital-diagonal hybridization~\cite{BauernfeindZinglTrieblEtAl2017}. Tensor-network influence-functional methods compress temporal correlations, but extending current single-impurity~\cite{ParkNgReichmanChan2024,SonnerLinkAbanin2025} and two-impurity Anderson implementations~\cite{LotemSonnerLinkEtAl2026} to general multiorbital models with off-diagonal hybridizations remains challenging.

These limitations motivate a deterministic approach to general multiorbital impurities with direct access to real-time Green's functions. In this Letter, we introduce Coupled-Cluster Coupled Lindblad (C$^3$L), a framework rooted in compact bath representations and quantum-chemistry perspectives on many-body correlations. Bath discretization maps a continuum environment onto a finite set of auxiliary modes, enabling explicit many-body treatments of the resulting system. However, unitary and independent Lorentzian representations require bath sizes that grow polynomially with the target time $T$~\cite{PriorChinHuelgaEtAl2010,DeVegaSchollwockWolf2015,TrivediMalzCirac2021,HuangParkChanLin2026}. C$^3$L instead employs the recently developed  coupled-Lindblad representation of non-Markovian baths~\cite{HuangParkChanLin2026,HuangLinXie2026}, which reliably reproduces the bath hybridization to accuracy $\varepsilon$ up to  time $T$ using only $\operatorname{polylog}(T/\varepsilon)$ coherently and dissipatively coupled pseudomodes, while maintaining numerically stable dynamics~\cite{ParkHuangZhuEtAl2024}.

Nevertheless, efficiently computing the steady state and Green's functions of the resulting interacting Lindbladian is still demanding. For instance, general impurity interactions and Lindbladian couplings would complicate real-time tensor-network calculations~\cite{WeimerKshetrimayumOrus2021,SulzLubichCerutiEtAl2024,WesthoffMoroderSchollwockPaeckel2025}.
A mixed-configuration approximation within the auxiliary master equation approach (AMEA) seeks to address this by averaging conditional single-orbital solutions, which could limit its description of complex interimpurity quantum fluctuations~\cite{MazzocchiTommasoWernerEtAl2025,MazzocchiAichhornArrigoni2026}.
Coupled-cluster (CC) theory provides an efficient route by explicitly including general interorbital interactions and organizing correlations into a hierarchy of excitations. Its singles-and-doubles truncation (CCSD) is known to be accurate for weak to intermediate correlations in molecules and solids~\cite{BartlettMusial2007,HirataPodeszwaTobitaBartlett2004,GruneisBoothMarsmanEtAl2011,BoothGruneisKresseAlavi2013,McClainSunChanBerkelbach2017,GruberLiaoTsatsoulisEtAl2018}. Its equation-of-motion (EOM) formulation provides direct access to charged and neutral excitations and Green's functions~\cite{StantonBartlett1993,KochJorgensen1990,McClainLischnerWatsonEtAl2016,PengKowalski2018,SheeZgid2019,ZhuJimenezHoyosMcClainEtAl2019,CoveneyTew2026}.

Extending CC and EOM-CC from Hamiltonians to general Lindbladians, beyond the specific settings treated previously~\cite{DzhioevKosov2015,ShushkovMiller2019,WernerLotzeArrigoni2023,MazzocchiTommasoWernerEtAl2025}, requires nontrivial theoretical machinery. We formulate a new parity-resolved fermionic superoperator tensor algebra on a doubled Fock space for steady states and Green's functions~\cite{CirioKuoChenEtAl2022,HuangZhuParkLin2025,BertrandBesserveFerreroEtAl2025}. A matrix-free Newton--Krylov solver yields the CC steady state, and Arnoldi-based EOM-CC yields real-time and real-frequency response functions without sampling or analytic continuation.  A two-orbital Anderson-impurity benchmark that took inchworm QMC tens of thousands of CPU hours~\cite{ErpenbeckBlommelZhangEtAl2024} now takes our prototype C$^3$L implementation $241$ CPU seconds, or $20$ seconds with GPU acceleration. We also solve a biased four-orbital impurity with general interactions and bath hybridizations, resolving transport resonances relevant to single-molecule scanning tunneling microscopy (STM) experiments.

\emph{Problem setup.}---We consider $n_{\rm I}$ interacting spinful impurity orbitals with Hamiltonian
\begin{equation}\label{eq:impurity_hamiltonian}
\hat H_{\rm I}
=\sum_{ij\sigma}h_{ij}\hat a_{i\sigma}^{\dagger}\hat a_{j\sigma}
+\sum_{ijkl,\sigma\sigma'}V_{ijkl}
\hat a_{i\sigma}^{\dagger}\hat a_{j\sigma'}^{\dagger}
\hat a_{k\sigma'}\hat a_{l\sigma}.
\end{equation}
For a Gaussian bath, the lesser and greater hybridization functions $\Delta^{\gtrless}(t)$ determine its full influence on impurity dynamics and correlation functions~\cite{CirioKuoChenEtAl2022,HuangZhuParkLin2025}. The coupled-Lindblad construction reconstructs the bath with accuracy $\varepsilon$ over a time interval of length $T$ using only $ N_{\rm pm} = \operatorname{polylog}(T/\varepsilon)$ modes through a numerically robust fitting procedure~\cite{HuangParkChanLin2026}, providing a compact representation scheme of non-Markovian impurity dynamics.
The impurity problem thus reduces to computing the steady state $\rho_{\rm ss}$ and the retarded Green's function $G^R_{i\sigma,j\sigma'}(t)=-\I\Theta(t)\operatorname{Tr}\!\left(\{\hat a_{i\sigma}(t),\hat a_{j\sigma'}^\dagger\}\rho_{\rm ss}\right)$
 for the Lindblad equation $\dv{\rho}{t}=\mc L[\rho]=-\I[\hat H_{\rm CL},\rho]+\mc D[\rho]$ defined by the coupled-Lindblad representation (see End Matter). Spectra and transport properties follow from these Green's functions. Other Green's-function components are treated similarly.

\begin{figure}[h]
  \centering
  \includegraphics[width=0.85\columnwidth]{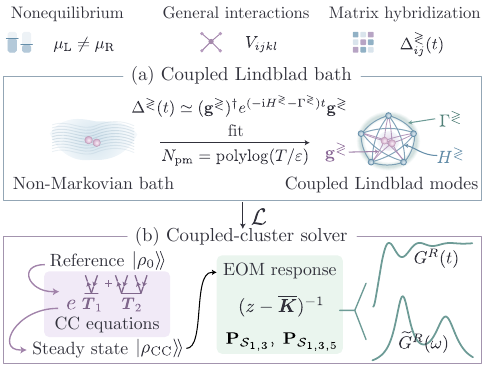}
  \caption{Overview of the C$^3$L solver. (a) Coupled-Lindblad pseudomodes reproduce the bath hybridization with only $\operatorname{polylog}(T/\varepsilon)$ modes~\cite{HuangParkChanLin2026}. (b) CC parametrizes the correlated steady state. The Green's functions are then computed using the EOM-CC formalism directly in the frequency domain.}
  \label{fig:method-overview}
\end{figure}

\begin{figure*}[t]
\centering
\begin{subfigure}{\textwidth}
\includegraphics[width=0.88\textwidth]{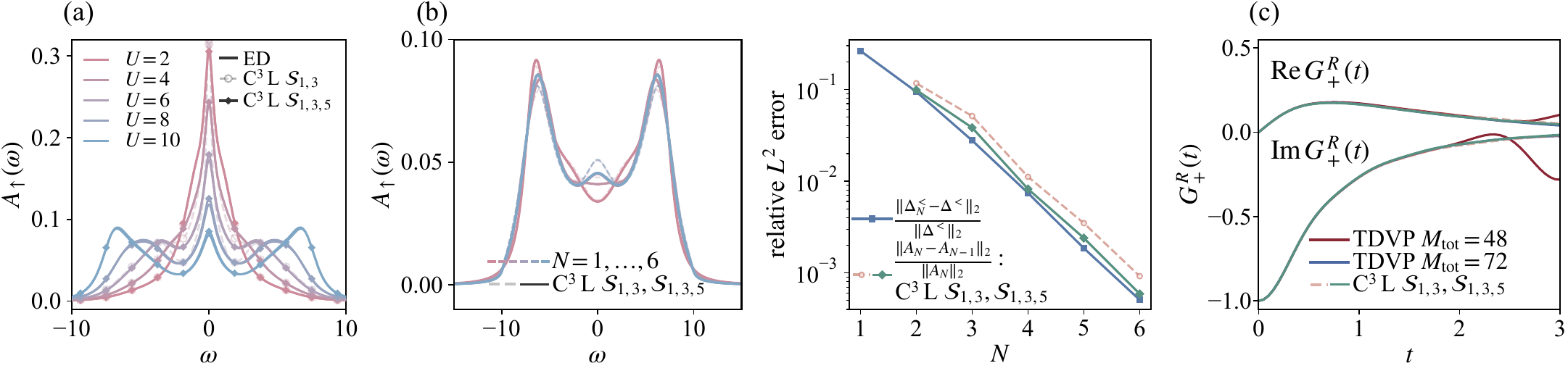}
\phantomsubcaption\label{fig:monomer-spectra}
\phantomsubcaption\label{fig:siam-u10}
\phantomsubcaption\label{fig:dimer}
\end{subfigure}
\caption{(a) Single-impurity spectra for $U=2,4,6,8,10$, $N_{\rm pm}=2$, $\beta=100$, compared with ED. (b) At $U=10$ and $\beta=1$, spectral functions for $N_{\rm pm}=1,\ldots,6$ (left), and the exponential convergence of both hybridization functions and spectra (right). (c) The bonding response $G_+^R(t) =G_{11}^R+G_{12}^R$ for the two-impurity Anderson model at $U=2$ and $\beta=\infty$, with $M_{\rm tot}=48,72$ unitary-bath spin orbitals. The corresponding hybridization fits are shown in Supplemental Material Fig.~\ref{fig:dimer-bath-fit}.}
\label{fig:hubbard-benchmarks}
\end{figure*}

\emph{Coupled-cluster steady state and response.}---With the bath reduced to a few pseudomodes, it remains to compute the steady state and Green's functions of the interacting Lindbladian.
CC achieves polynomial cost with good accuracy at intermediate correlation by
parametrizing the correlated steady state in terms of excitation amplitudes $\bv t$ through
 $
\kett{\rho_{\rm CC}}=e^{\bs T(\bv t)}\kett{\rho_0},
 $ where $\kett{\bullet}$ denotes the vectorization of a density matrix. In CCSD, $\bs T$ contains single and double excitations about the Gaussian reference $\rho_0$.
EOM-CC then uses the converged steady-state amplitudes and iterative linear solvers to compute the frequency-domain Green's function. In this work, we truncate the EOM space to $\mc S_{1,3}$ or $\mc S_{1,3,5}$, containing configurations of ranks $1,3$ or $1,3,5$, respectively.

C$^3$L (\cref{fig:method-overview}) addresses two distinct computational costs by reducing the number of orbitals needed to represent the bath, and using low-scaling solvers for the resulting correlated problem. We note that real-time QMC methods that treat the bath continuously can still entail costs growing exponentially with impurity size $n_{\rm I}$, and linearly with the time window $T$ \footnote{For example, the general multiorbital inchworm implementation uses dense impurity-space matrix products costing $\mc O((4^{n_{\rm I}})^3)$, for $n_{\rm I}$ spinful impurity orbitals. Its steady-state formulation scales linearly with the number of relative-time grid points $T/\Delta t$ at fixed sampling effort, expansion order, and iteration count~\cite[Sec.~II.C]{ErpenbeckBlommelZhangEtAl2024}.}.
By comparison, C$^3$L achieves polynomial scaling in $n_{\rm I}$ and polylogarithmic scaling in $T$. As a deterministic solver, it also avoids the potentially large prefactor associated with stochastic sampling. Methodological details are provided in the End Matter and Supplemental Material Secs.~\ref{sec:CCSD} and \ref{sec:EOM-CC}.

\emph{Single- and two-impurity Anderson models.}---%
We first benchmark systems accessible to ED and time-dependent variational principle (TDVP) calculations, starting with the particle--hole-symmetric single-impurity Anderson model with a semicircular bath. Numerical setups and additional results are given in Supplemental Material Sec.~\ref{sec:numerical_setups}.
In \cref{fig:monomer-spectra} ($\beta = 100$, $N_{\rm pm } = 2$), the $\mc S_{1,3}$ response agrees well with the ED spectra $A(\omega)=-\frac{1}{\pi}\Im\widetilde G^R(\omega)$ of the same coupled-Lindblad problem at $U=2$. Including $\abs{S}=5$ excitations maintains close agreement at larger $U$, up to $U = 10$. This is consistent with Hamiltonian Green's-function CC (GFCC) results: the CCSD state remains accurate at intermediate interaction strengths, while the Green's function requires higher EOM ranks as interactions strengthen \cite{ZhuChan2021,SheeYehPengEtAl2023}. At $U=10$ and $\beta=1$, \cref{fig:siam-u10} shows that both the hybridization fit and the spectra converge exponentially with the number of pseudomodes, illustrating the compactness of the coupled-Lindblad representation~\cite{HuangZhuParkLin2025,HuangParkChanLin2026}.

We next benchmark a two-impurity Anderson model against TDVP propagation with a discretized unitary bath~\cite{DeVegaSchollwockWolf2015,PriorChinHuelgaEtAl2010}. Each impurity couples to an independent semicircular bath at $\beta = \infty$. For a fixed time step, the number of sequential TDVP propagation steps grows linearly with the target time. C$^3$L instead solves directly in the real-frequency domain, allowing $G^R(t)$ over the fitted interval $T$ to be reconstructed by a single inverse Fourier transform. In \cref{fig:dimer}, the TDVP bonding response $G_+^R$ loses accuracy at $t\approx2$ for $M_{\rm tot}=48$ unitary-bath spin orbitals. Increasing the unitary bath to $M_{\rm tot}=72$ restores agreement with C$^3$L up to $t=3$, whereas the coupled-Lindblad representation uses only $N_{\rm pm}=4$ pseudomodes. C$^3$L thus benefits from both a smaller bath and the low-scaling cost of the CC solver.

\begin{figure*}[t]
\centering
\includegraphics[width=0.95\textwidth]{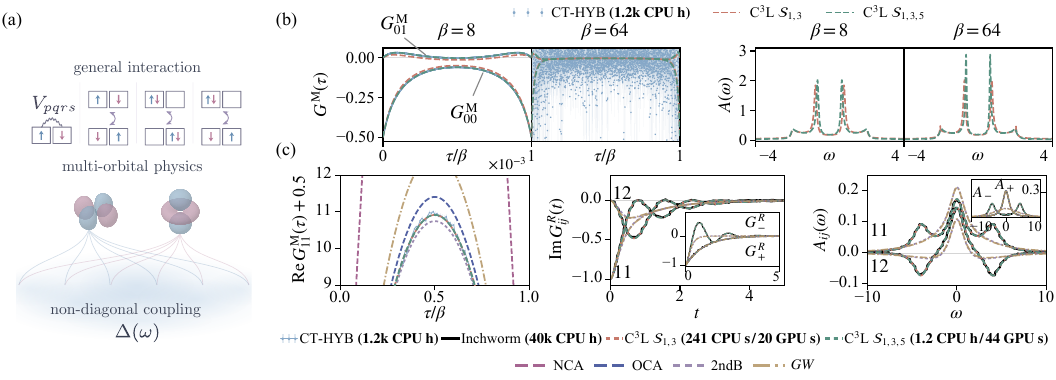}
{\phantomsubcaption\label{fig:multiorbital-schematic}
\phantomsubcaption\label{fig:kanamori-equilibrium}
\phantomsubcaption\label{fig:general-multiorbital}}
\caption{(a) Schematic of general interorbital interactions and off-diagonal hybridization. (b) Kanamori model at $\beta = 8$ and $\beta = 64$ \cite{EidelsteinGullCohen2020}: $G^{\rm M}(\tau)$ and the C$^3$L spectra $A(\omega)$ per spin. Error bars on CT-HYB points indicate one standard error. The average sign is only $8.32\times10^{-4}$ at $\beta = 64$. (c) General interacting two-orbital model at $\beta=0.1$ with the inchworm result digitized from Ref.~\cite{ErpenbeckBlommelZhangEtAl2024}: imaginary-time Green's functions (left); $\Im G_{ij}^R(t)$ (middle) and $A_{ij}(\omega)$ (right). Insets resolve $\Im G^R_\pm$ and $A_\pm$ in the $\pm $ channels, where $G^R_\pm=\frac{G^R_{11}+G^R_{22}\pm G^R_{12}\pm G^R_{21}}{2}$ and likewise for $A_\pm$.  }
\label{fig:multiorbital-benchmarks}
\end{figure*}

\emph{General interactions in multiorbital impurities.}---We compare with imaginary- and real-time QMC solvers for multiorbital problems with general interactions and bath hybridizations (\cref{fig:multiorbital-schematic}).
We first consider the rotationally invariant two-orbital Kanamori interaction, relevant to two $e_{\rm g }$ bands in correlated transition metals~\cite{Kanamori1963,EidelsteinGullCohen2020},
\begin{equation} \label{eq:kanamori-hamiltonian}
\begin{aligned}
\hat H_{\rm I}& =U\sum_i \hat n_{i\downarrow}\hat n_{i\uparrow}
+\sum_{i<j,\sigma\sigma'}(U'-J^{\rm H}\delta_{\sigma\sigma'})\hat n_{i\sigma}\hat n_{j\sigma'} \\
&+J^{\rm H}\sum_{i\ne j}\left(\hat a_{i\uparrow}^\dagger \hat a_{j\downarrow}^\dagger
\hat a_{i\downarrow}\hat a_{j\uparrow}+\hat a_{i\uparrow}^\dag \hat a_{i\downarrow}^\dag \hat a_{j\downarrow}\hat a_{j\uparrow}\right).%
\end{aligned}
\end{equation}
We consider the half-filled model with $U=2$, $J^{\rm H}=0.2$, and $U'=U-2J^{\rm H}$, coupled to a semicircular bath with equal diagonal and off-diagonal couplings. Additional benchmarks away from half filling are presented in Supplemental Material Sec.~\ref{sec:numerical_setups}.
We compute the imaginary-time Green's function from the C$^3$L solution at Matsubara frequencies.
Our CT-HYB calculations use the \texttt{TRIQS} library \cite{ParcolletFerreroAyralEtAl2015} and 1,200 CPU hours per temperature.
Although CT-HYB generally provides numerically exact imaginary-time Green's functions, Kanamori interactions and off-diagonal bath couplings cause a severe  low-temperature sign problem~\cite{EidelsteinGullCohen2020,KayeHuangStrandGolez2024}. As shown in \cref{fig:kanamori-equilibrium}, C$^3$L with $\mc S_{1,3,5}$ agrees well with CT-HYB at $\beta=8$ and remains stable at $\beta=64$, where CT-HYB becomes extremely noisy. C$^3$L also provides real-frequency spectra directly, without analytic continuation.

 The second two-orbital case tests the solver directly against state-of-the-art real-time inchworm QMC, using the general interacting model of Ref.~\cite{ErpenbeckBlommelZhangEtAl2024}:
\begin{equation}
\begin{aligned}
 \hat H_{\rm I}=&{}\sum_{i\sigma}[\epsilon_0\hat n_{i\sigma} +V \hat n_{i\sigma}
(\hat a_{1\bar\sigma}^\dagger \hat a_{2\bar\sigma}+\mathrm{h.c.})]+U\sum_{ij}\hat n_{i\uparrow}\hat n_{j\downarrow}
\\&
+\sum_\sigma[\nu(\hat a_{1\sigma}^\dagger \hat a_{2\sigma}+\mathrm{h.c.})+ V' \hat a_{1\sigma}^\dagger \hat a_{1\bar\sigma}
\hat a_{2\bar\sigma}^\dagger \hat a_{2\sigma} ] \\& +\nu'(\hat a_{1\uparrow}^\dagger \hat a_{1\downarrow}^\dagger
\hat a_{2\uparrow}\hat a_{2\downarrow}+\mathrm{h.c.}).
\end{aligned}
\label{eq:inchmoss-benchmark}
\end{equation}
The parameters and the off-diagonal wide-band baths are specified in Supplemental Material Sec.~\ref{sec:numerical_setups}. We compare C$^3$L with imaginary-time hybridization-expansion solvers in the non-crossing (NCA) and one-crossing (OCA) approximations~\cite{KayeHuangStrandGolez2024}, and with real-time solvers based on fully self-consistent 2ndB and $GW$ MBPTs. CT-HYB and inchworm QMC provide the respective numerically exact benchmarks.
As shown in \cref{fig:general-multiorbital}, C$^3$L gives the closest agreement with the 1,200-CPU-hour CT-HYB imaginary-time Green's function. In real time, C$^3$L at both EOM ranks reproduces the order-$12$ inchworm QMC results, whereas 2ndB and $GW$ do not. The inchworm QMC calculation requires tens of thousands of CPU hours, whereas the combined C$^3$L calculations of imaginary- and real-time Green's functions require only $241$ CPU seconds, or $20$ seconds on a single NVIDIA H200 GPU, with $\mc S_{1,3}$, and $1.2$ CPU hours or $44$ GPU seconds with $\mc S_{1,3,5}$.

\emph{Resonance-resolved transport in a molecular junction.}---A tunneling electron can leave a molecule in different many-body states even when it exits through the same orbital. In STM experiments on perylenetetracarboxylic dianhydride (PTCDA), bias-dependent conductance and light emission reveal selective access to neutral triplet excitons through electron removal from the anion~\cite{KimuraMiwaImadaEtAl2019}. Controlled single-electron spectroscopy further distinguishes transitions between molecular charge states and neutral excitations~\cite{SelliesEckrichGrossEtAl2025}. When an extra electron occupies the lowest unoccupied molecular orbital (LUMO), removing an electron from the highest occupied molecular orbital (HOMO) leaves two unpaired electrons. Their spins can form a neutral triplet or excited singlet, $\mathbf D^-\to\mathbf T,\mathbf S_1$. Spin exchange lowers the triplet relative to the corresponding singlet, allowing the triplet-related transition to enter transport at a smaller bias voltage. Removing a HOMO electron from the closed-shell neutral molecule instead leaves a cationic doublet, $\mathbf S_0\to\mathbf D^+$.

\begin{figure*}[t]
\centering
\captionsetup[subfigure]{labelfont=normal}
\begin{subfigure}[t]{.34\textwidth}
\vspace{0pt}\phantomsubcaption\label{fig:stm-device}
\makebox[\linewidth][l]{{\fontfamily{ptm}\selectfont\fontsize{8}{9}\selectfont(a)}}\par
\includegraphics[width=\linewidth]{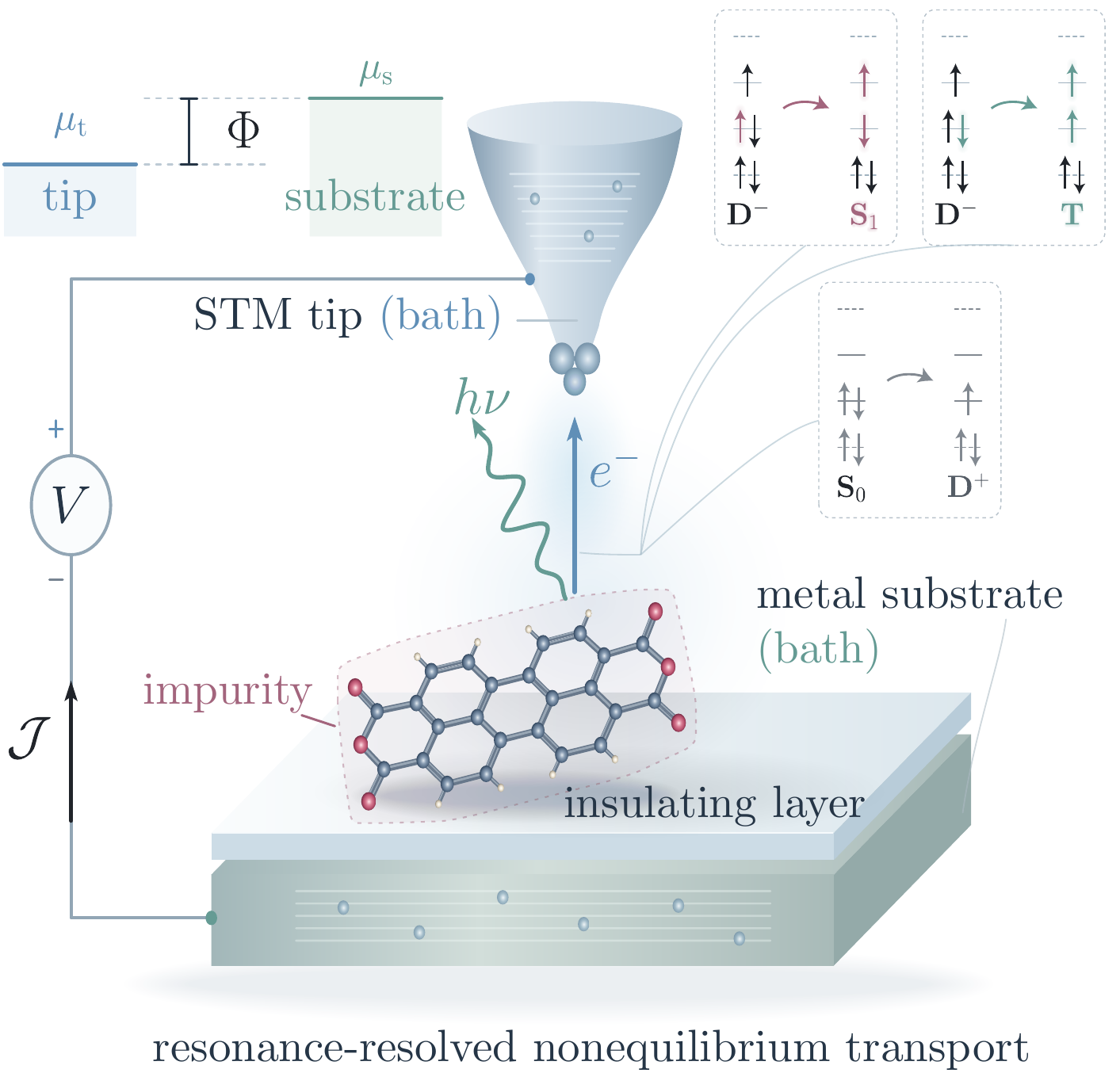}
\end{subfigure}\hfill
\begin{minipage}[t]{.61\textwidth}
\vspace{0pt}
\begin{subfigure}[t]{.5\linewidth}
\vspace{0pt}\phantomsubcaption\label{fig:stm-lumo-matsubara}
\makebox[\linewidth][l]{{\fontfamily{ptm}\selectfont\fontsize{8}{9}\selectfont(b)}}\par
\includegraphics[width=\linewidth]{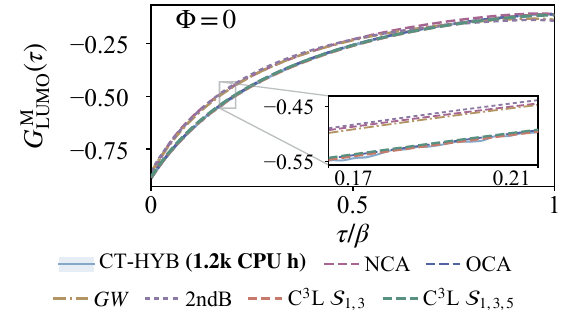}
\end{subfigure}\hfill
\begin{subfigure}[t]{.495\linewidth}
\vspace{0pt}\phantomsubcaption\label{fig:stm-neutral-ratio}
\makebox[\linewidth][l]{{\fontfamily{ptm}\selectfont\fontsize{8}{9}\selectfont(c)}}\par
\includegraphics[width=\linewidth]{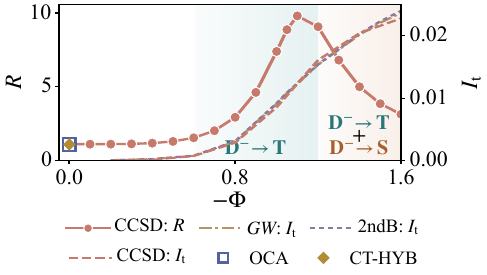}
\end{subfigure}
\par\smallskip
\includegraphics[width=\linewidth]{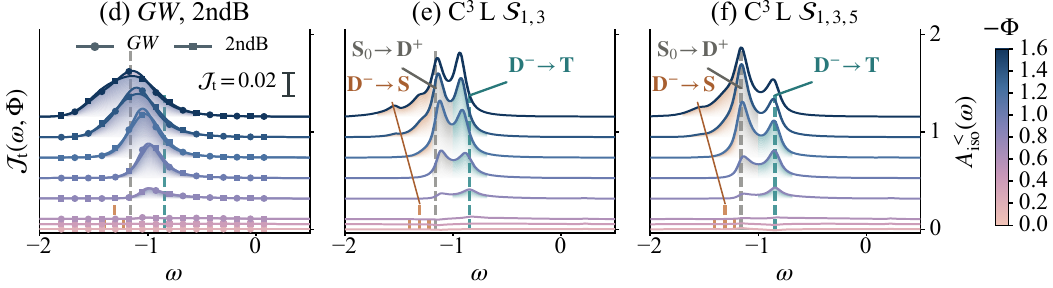}
{\phantomsubcaption\label{fig:stm-mbpt-current}
\phantomsubcaption\label{fig:stm-s13-current}
\phantomsubcaption\label{fig:stm-s135-current}}
\end{minipage}
\caption{Resonance-resolved molecular transport.
(a) Frontier orbitals coupled to tip and substrate reservoirs, with charge-extraction transitions \cite{KimuraMiwaImadaEtAl2019}.
(b) Equilibrium $G_{\rm LUMO}^{\rm M}(\tau)$.
(c) CCSD neutral-strength ratio $R$ and tip currents $I_{\rm t}$. Markers denote equilibrium OCA and CT-HYB ratios at $\Phi = 0$.
(d--f) Extraction current $\mathcal J_{\rm t}$ for $GW$/2ndB and C$^3$L. Dashed sticks show isolated-molecule removal spectra, evaluated for the initial neutral singlet or anionic doublet. Shaded regions denote the candidate assignments. The CT-HYB cost in (b) refers to the equilibrium calculation ($\Phi = 0$). }
\label{fig:stm-exchange-transport}
\end{figure*}

We model these competing removal processes with four spinful frontier orbitals $(\mathrm{h}_2,\mathrm{h}_1,\mathrm{l}_1,\mathrm{l}_2)$ coupled to metallic tip and substrate baths through off-diagonal hybridization (\cref{fig:stm-device}), with tip bias $\Phi$. The impurity Hamiltonian is
\begin{equation}\label{eq:stm-exchange-hamiltonian}
\begin{aligned}
\hat H_{\rm I}={}&\sum_{ij\sigma}h_{ij}\hat a_{i\sigma}^\dagger\hat a_{j\sigma}
+U\sum_i\hat n_{i\uparrow}\hat n_{i\downarrow}
+V\sum_{i<j}\hat n_i\hat n_j\\
&-\sum_{a,b=1}^2 J^{\rm E}_{\mathrm{h}_a \mathrm{l}_b}
\left(2\hat{\bv S}_{\mathrm{h}_a}\cdot\hat{\bv S}_{\mathrm{l}_b}
+\frac12\hat n_{\mathrm{h}_a}\hat n_{\mathrm{l}_b}\right)\\
&-\sum_{a,b=1}^2 J^{\rm P}_{\mathrm{h}_a \mathrm{l}_b}
\left(\hat a_{\mathrm{h}_a\uparrow}^\dagger\hat a_{\mathrm{h}_a\downarrow}^\dagger
\hat a_{\mathrm{l}_b\downarrow}\hat a_{\mathrm{l}_b\uparrow}+\mathrm{h.c.}\right).
\end{aligned}
\end{equation}
Here $\hat n_i:= \hat n_{i\uparrow} + \hat n_{i\downarrow}$ and $\hat{\bv S}_i$ are the orbital occupation and spin. $J^{\rm E}$ and $J^{\rm P}$ describe exchange and pair hopping. The additional orbitals allow competing configurations beyond the  HOMO--LUMO pair. This effective electronic impurity model is motivated by the experiments, and its parameters are given in Supplemental Material Sec.~\ref{sec:numerical_setups}.

We first benchmark the equilibrium response.
At $\Phi=0$ and $\beta=10$, the C$^3$L imaginary-time LUMO Green's function agrees well with CT-HYB and OCA (\cref{fig:stm-lumo-matsubara}).
Increasing the extraction bias $-\Phi$ raises the tip current $I_{\rm t}$, with similar trends in CCSD, $GW$, and 2ndB (\cref{fig:stm-neutral-ratio}). The driven steady state's neutral spin response changes markedly as $-\Phi$ increases. We characterize it by $R$, the ratio of the spin-vector orbital-transfer strength per component to the scalar strength~\cite{KanekoOhta2014,PokhilkoKrylov2019}. For the active orbital pair, $R>1$ ($R<1$) indicates triplet-like (singlet-like) correlations. The rise and subsequent decline of $R$ are consistent with preferential access to the lower-bias triplet-related resonance followed by increasing excited-singlet contributions. This ordering agrees with the exchange-selected sequence observed experimentally~\cite{KimuraMiwaImadaEtAl2019}.

Although CCSD, $GW$, and 2ndB predict similar total currents (\cref{fig:stm-neutral-ratio}), the energy distributions of the tip current are different, as revealed by the nonequilibrium energy-resolved current $\mc J_{\rm t}(\omega,\Phi)$. We compute $\mc J_{\rm t}(\omega,\Phi)$ from the molecular removal and addition spectral functions $A^{\gtrless}(\omega,\Phi)$ using the Meir--Wingreen formula~\cite{MeirWingreen1992,JauhoWingreenMeir1994}. In \cref{fig:stm-mbpt-current,fig:stm-s13-current,fig:stm-s135-current}, $GW$ and 2ndB produce broad envelopes, whereas C$^3$L resolves distinct resonances whose relative weights vary with bias.
Isolated-molecule removal spectra, evaluated with the tip-coupling operator, suggest contributions from anionic triplet-producing removal and ordinary neutral ionization around $\Phi=-0.6$. At larger extraction bias, the excited-singlet family contributes to the lower-energy shoulder. The resolved transport features thus follow the same sequence of removal processes as in experiment.
Enlarging the EOM space from $\mc S_{1,3}$ to $\mc S_{1,3,5}$ preserves the resonance-resolved behavior, although individual peak shapes change. The greater resolution of many-body structure in C$^3$L is consistent with results for GFCC impurity spectra~\cite{ZhuChan2021}, electron-gas satellites~\cite{McClainLischnerWatsonEtAl2016}, and X-ray absorption features~\cite{RehrVilaKasEtAl2020}.

Exchange interactions produce multiplet structure in the removal spectrum, whose resolution requires correlating the particle--hole pairs created by electron removal. To illustrate a possible failure mechanism, we consider a minimal exchange model in Supplemental Material \cref{app:multiplet_GF_mbpt}. Although the 2ndB and $GW$ approximations include exchange coupling, they do not fully correlate the intermediate particle--hole configurations. This introduces a spurious charging energy that suppresses exchange-induced mixing, so that distinct multiplet peaks collapse into a single dominant resonance.

\emph{Conclusions and outlook.}---C$^3$L is an efficient and accurate solver for general multiorbital impurity models both in and out of equilibrium. Already at two orbitals, it accesses regimes where quantum Monte Carlo methods become prohibitively expensive for routine calculations. It further resolves four-orbital finite-bias transport in molecular junctions, well beyond what CT-QMC, inchworm, and influence-functional solvers have demonstrated.

Robust nonlinear solvers~\cite{MudiyanselageLiFaulstich2026} and Krylov methods tailored to nonnormal operators~\cite{GreenbaumPtakStrakos1996} could broaden the regimes accessible to C$^3$L. The method can also be used in nonequilibrium dynamical mean-field theory (DMFT) calculations~\cite{SavrasovHauleOudovenkoEtAl2006,AokiTsujiEckstein2014}. Connections to diagrammatic many-body theory for Lindbladian Green's functions remain to be explored~\cite{BlommelPerfettoStefanucciVlcek2026,LangeBerkelbach2018,TolleKitsarasIrmlerEtAl2026}.

Extending C$^3$L to bosonic baths would allow applications to electron--phonon coupled systems \cite{WhiteGaoMinnichChan2020} and trapped-ion systems \cite{LemmerCormickTamascelliEtAl2018,KangNuominYulyEtAl2024}. C$^3$L could also provide classical benchmarks for system--bath algorithms on early fault-tolerant quantum hardware~\cite{DingZhanPreskillLin2026,WangDing2025,LiYangLin2026}.

\emph{Data and Code Availability.}---All C$^3$L benchmarks and simulations were conducted using the  \texttt{C3L} package publicly available on GitHub \cite{c3l}.

\emph{Acknowledgments.}---
This material is based upon work supported by the U.S. Department of Energy, Office of Science, Accelerated Research in Quantum Computing Centers, Quantum Utility through Advanced Computational Quantum Algorithms, grant no. DE-SC0025572 (H.L.), and by the Simons Targeted Grants in Mathematics and Physical Sciences on Moir\'e Materials Magic (Z.H.). H.L. thanks the Department of Computing and Mathematical Sciences at Caltech, where part of this work was done, for its hospitality. The Flatiron Institute is a division of the Simons Foundation. L.L. is a Simons Investigator in Mathematics. We thank Andre Erpenbeck, Antoine Georges, Emanuel Gull, Andy Millis, Gunhee Park, Xiao-Yu Ouyang, and Avijit Shee for helpful discussions and suggestions.

OpenAI GPT 5.6 Sol and 6 Astra were used to assist with code development and algebraic derivations under the authors’ direction and careful examination. We estimate that completing the same work without this assistance would have taken years. We take full responsibility for the content of this work.

\nocite{Kalman1965}
\bibliography{ref}
\onecolumngrid
\section*{End Matter}
\label{sec:endmatter}

\twocolumngrid

\emph{Coupled-Lindblad bath representation.}---In the coupled-Lindblad construction, we introduce a finite set of auxiliary pseudomodes by fitting the greater and lesser hybridization functions as
\begin{equation}
\Delta^{\gtrless}(t)\simeq
(\bv g^{\gtrless})^\dagger e^{(-\I H^{\gtrless}-\Gamma^{\gtrless})t}\bv g^{\gtrless}
\end{equation}
where $\bv g^{\gtrless} \in \mathbb C^{N_{\rm pm}^{\gtrless} \times n_{\rm I}}$ and Hermitian $H^{\gtrless},\Gamma^{\gtrless}\in\mathbb C^{N_{\rm pm}^\gtrless \times N_{\rm pm}^{\gtrless}}
$, with $\Gamma^{\gtrless} \succeq 0$. In practice, we typically choose $N_{\rm pm}  =N_{\rm pm}^> = N_{\rm pm}^<$. The enlarged coupled-Lindblad Hamiltonian is
\begin{equation}\label{eq:coupled-lindblad-hamiltonian}
\begin{aligned}
\hat H_{\rm CL}&={}\hat H_{\rm I}
+\sum_{k,l,\sigma}(H^<_{kl}\hat p_{k\sigma}^\dagger\hat p_{l\sigma}
+ H^>_{kl}\hat d_{k\sigma}^\dagger\hat d_{l\sigma})\\
&+\sum_{i,k,\sigma}\left[
(\bv g^<)_{ki}\hat p_{k\sigma}^\dagger\hat a_{i\sigma}
+(\bv g^>)_{ki}\hat d_{k\sigma}^\dagger\hat a_{i\sigma}
+\mathrm{h.c.}\right].
\end{aligned}
\end{equation}
Here $\hat p_{k\sigma}$ ($\hat d_{k\sigma}$) annihilate the lesser (greater) pseudomodes. The coupled-Lindblad dissipator is
\begin{equation}\label{eq:dissipator}
    \begin{aligned}
        \mc D[\rho ] =& \sum_{k,l,\sigma} \Gamma^{<}_{kl}   (2\hat p_{k\sigma}^\dagger \rho \hat p_{l\sigma} -  \{\hat p_{l\sigma} \hat p_{k\sigma}^\dagger,\rho\})\\
      & +  \sum_{k,l,\sigma} \Gamma^{>}_{kl}  ( 2\hat d_{l\sigma} \rho \hat d_{k\sigma}^\dagger - \{\hat d_{k\sigma}^\dagger \hat d_{l\sigma},\rho\} ).
    \end{aligned}
\end{equation}
Together, the Hamiltonian and dissipator define the Lindblad dynamics $\dv{\rho}{t}=\mc L[\rho]=-\I[\hat H_{\rm CL},\rho]+\mc D[\rho]$ on the enlarged space. For $d$ independent bath channels, the enlarged system contains at most $n=n_{\rm I}+2d N_{\rm pm}$ spinful orbitals \footnote{More precisely, consider separable hybridizations $\Delta^{\gtrless}(t)=\sum_{i=1}^{d}c_i^{\gtrless}(t)\bv u_i\bv u_i^\dagger$, with linearly independent $\bv u_i$. If the Laplace transforms of the scalar functions $c_i^{\gtrless}$ are rational (or rationally approximated) with $N_{{\rm pm},i}^{\gtrless}$ poles \cite{Kalman1965}, any exact realization of these rational functions requires at least $\sum_i N_{{\rm pm},i}^{\gtrless}$ auxiliary modes. This bound is attained by independent bath channels whenever each scalar function admits a Lindblad realization of the corresponding degree.}. If the hybridization function is fitted to accuracy $\varepsilon$ over $[0,T]$, the pseudomode dynamics reproduces the impurity reduced density operator $\rho_{\rm I}(t)$ for any $t\in [0,T]$, with an error that vanishes as $\varepsilon\to0$~\cite{HuangZhuParkLin2025}. The coupled-Lindblad construction requires only $N_{\rm pm} = \mathrm{polylog}(T/\varepsilon)$ pseudomodes, providing a compact representation for long-time dynamics.

The fitting problem admits a convex formulation that can be solved robustly and efficiently using semidefinite programming (SDP), avoiding the nonconvex optimization required by many existing approaches, which can suffer from numerical instabilities in practice~\cite{ArrigoniKnapVonDerLinden2013,WernerLotzeArrigoni2023}. The procedure is implemented in the publicly available \texttt{RealTimeBath} package on GitHub~\cite{RealTimeBath}.

\emph{Coupled cluster equations.}---%
We vectorize $\rho$ as $\kett{\rho}$ and set $\bs K=\I\bs L$, where $\bs L\kett{\rho} := \kett{\mathcal L[\rho]}$.  Let $\bs c_p$, $\wt{\bs c}_p$, and their adjoints denote the fermionic superoperators. A self-consistent Hartree--Fock solution provides a Gaussian reference $\kett{\rho_0}$ and biorthogonal normal-mode superoperators $\{\bs P_\kappa,\widetilde{\bs P}_\kappa\}$. We partition the Lindbladian as $
    \bs K=\bs K_0+\bs W,$ where
    \begin{equation}
    \bs K_0=\sum_\kappa\lambda_\kappa\widetilde{\bs P}_\kappa\bs P_\kappa\quad  \text{satisfies}  \quad \bs K_0\kett{\rho_0} = 0,
    \end{equation}
 and $\bs W$ is an even polynomial in $\{\widetilde{\bs P}_\kappa,  {\bs P}_\kappa \}$ of degree at most $4$. We define the configuration $\kett{\Phi_S}=\bs\tau_S\kett{\rho_0}$, where $\bs\tau_S$ is the product of normal-mode creators $\widetilde{\bs P}_\kappa$ with $\kappa\in S$.
 The symmetries of the Hamiltonian and dissipator restrict
  the steady state to the charge-neutral sector
 $  \mathrm{Span}\{\kett{\Phi_S}: Q_\uparrow(S) = Q_\downarrow(S) = 0\},
 $ as detailed in Supplemental Material Sec.~\ref{sec:HF}.

CC represents the correlated steady state $\kett{\rho_{\rm ss}}$ using the exponential ansatz $\kett{\rho_{\rm  CC }} := e^{\bs T}\kett{\rho_0}$~  \cite{BartlettMusial2007,ShavitBartlett2009}, with
\begin{equation}
       \bs T   := \sum_{r\geq 1}\bs T_r,\quad \bs T_r := \sum_{\substack{|S|=2r, Q_{\uparrow/\downarrow}(S)=0}} \mathrm{t}_S \boldsymbol \tau_S,
\end{equation}
  where
  $\bm{\mathrm{t}}:= \{\mathrm{t}_S\}$ denotes the set of complex amplitudes. In CCSD, we truncate the excitation rank $r$ at $2$ \begin{equation}
  S\in \mc S_{\rm CCSD} := \{\abs{S} = 2,4,Q_{\uparrow/\downarrow}(S) = 0\},
  \end{equation}requiring $\mc O(n^4)$ amplitudes for $n$ spinful orbitals~\cite{BartlettMusial2007,ShavitBartlett2009}. The CCSD amplitude equations are obtained by projecting $\bs K \kett{\rho_{\rm CC}} = 0$ onto the left configurations $\bbra{\wt \Phi_S}$
  \begin{equation}\label{eq:CCSD_amplitude_equation}
    z_S(\bm{\mathrm{t}}) := \bbra{\wt \Phi_S} e^{-\bs T (\bm{\mathrm{t}}) } \bs K e^{\bs T (\bm{\mathrm{t}}) }\kett{\rho_0} = 0,~\forall S\in \mc S_{\rm CCSD}.
  \end{equation}
  The residual can be evaluated at $\mc O(n^6)$ computational cost.
  Since each term in $\bs T$ is a nonempty creator string, $\bbra{I} \kett{\rho_{\rm CC}}    = 1$, so trace normalization holds automatically. Lindblad CC therefore requires neither the energy equation nor the intermediate normalization used in standard CC~\cite{ShavitBartlett2009}. The residual in \cref{eq:CCSD_amplitude_equation} and the Jacobian are evaluated through low-rank tensor contractions, enabling a matrix-free Newton--Krylov solver that improves nonlinear convergence~\cite{YangBrabecVeisEtAl2020,MudiyanselageLiFaulstich2026}, as detailed in Supplemental Material Sec.~\ref{sec:CCSD}.
We recall the retarded Green's function, with $p=(i,\sigma)$ labeling an impurity spin orbital,
\begin{equation}\label{eq:retarded-definition-endmatter}
G^R_{pq}(t)=-\I\Theta(t)\operatorname{Tr}\!\left[\{\hat a_p(t),\hat a_q^\dagger\}\rho_{\rm ss}\right].
\end{equation}
Since $\hat a_p$ is odd under particle-number parity, it evolves as $\hat a_p(t)=e^{t\underline{\mc L}^\dagger}[\hat a_p]$. Here $\underline{\mc L}^\dagger$ is the adjoint of the odd-parity generator
\begin{equation}
\underline{\mc L}[\bullet]=\hat\Pi\mc L[\hat\Pi\bullet], \quad
\hat\Pi=e^{\I\pi\hat N},
\end{equation}
where $\hat N$ is the total number operator~\cite{CirioKuoChenEtAl2022,HuangZhuParkLin2025,BertrandBesserveFerreroEtAl2025}. A Fourier transform of the projected dynamics yields the frequency-domain EOM-CC expression
\begin{equation}\label{eq:eom-resolvent-endmatter}
\begin{aligned}
\wt G_{pq}^R(z)={}&\bbra{I}\bs c_p  \bv P_{\mc S_{\rm EOM}}
(z  -\overline{\bs K} )^{-1} \bv P_{\mc S_{\rm EOM}}\bv P_{\rm o}
(\overline{\bs c_q^\dagger}-\I\overline{\wt{\bs c}_q})\kett{\rho_0}.
\end{aligned}
\end{equation}
In \cref{eq:eom-resolvent-endmatter}, $\bv P_{\rm o}=(\bs I-\bs\Pi)/2$, where $\bs\Pi$ is the parity superoperator. The response projector selects $\mc S_{1,3}$ or $\mc S_{1,3,5}$, with excitation ranks $|S|=1,3$ or $1,3,5$ and charge $+1$. At a converged CCSD solution, $\overline{\bs K} $ is the similarity-transformed superoperator $\overline {\bs K}: = e^{-\bs T(\bv t_{\rm CCSD})}\bs K e^{\bs T(\bv t_{\rm CCSD})}$. These spaces contain $\mc O(n^3)$ and $\mc O(n^5)$ configurations, respectively.

We also consider the lesser and greater Green's functions, defined for $t\geq0$ as
\begin{equation}\label{eq:lesser-greater-definitions}
\begin{aligned}
G_{pq}^<(t) =\I\operatorname{Tr}\!\left[\hat a_q^\dagger\hat a_p(t)\rho_{\rm ss}\right], \quad G_{pq}^>(t) =-\I\operatorname{Tr}\!\left[\hat a_p(t)\hat a_q^\dagger\rho_{\rm ss}\right].
\end{aligned}
\end{equation}
The CC calculation replaces $\rho_{\rm ss}$ by $\rho_{\rm CC}$. Splitting the source in \cref{eq:eom-resolvent-endmatter} gives the addition and removal contributions of $\wt G^R(z)$,
\begin{equation}\label{eq:lesser-greater-fourier}
\begin{aligned}
\wt G_{pq}^{R,>}(z)={}&\bbra{I}\bs c_p \bv P_{\mc S_{\rm EOM}}
(z -\overline{\bs K} )^{-1} \bv P_{\mc S_{\rm EOM}}\bv P_{\rm o}\overline{\bs c_q^\dagger}\kett{\rho_0},\\
\wt G_{pq}^{R,<}(z)={}&-\I\bbra{I}\bs c_p \bv P_{\mc S_{\rm EOM}}
(z - \overline{\bs K}  )^{-1} \bv P_{\mc S_{\rm EOM}}\bv P_{\rm o}\overline{\wt{\bs c}_q}\kett{\rho_0}.
\end{aligned}
\end{equation}
 On the real-frequency axis, the full greater and lesser functions can be evaluated as
\begin{equation}\label{eq:lesser-greater-from-components}
\begin{aligned}
\wt G^>(\omega)&=\wt G^{R,>}(\omega+\I0^+)-[\wt G^{R,>}(\omega+\I0^+)]^\dagger,\\
\wt G^<(\omega)&=-\wt G^{R,<}(\omega+\I0^+)+[\wt G^{R,<}(\omega+\I0^+)]^\dagger.
\end{aligned}
\end{equation}
We evaluate the similarity-transformed actions through tensor contractions and solve the resulting nonnormal linear systems using matrix-free Arnoldi-type iterative methods~\cite{SaadSchultz1986}. Shift invariance allows the same Krylov basis to be reused across frequencies. The detailed derivation is given in Supplemental Material Sec.~\ref{sec:EOM-CC}.

\begin{figure}[h]
\centering
\includegraphics[width=0.9\linewidth]{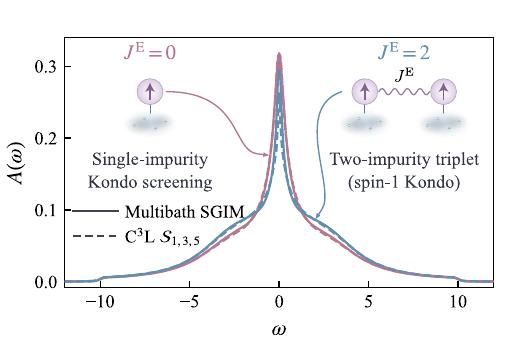}
\caption{Low-temperature Kondo spectra at $U=4$ and $\beta=100$ for $J^{\rm E}=0$ (red) and $2$ (blue). Solid curves: multibath SGIM data digitized from Ref.~\cite{LotemSonnerLinkEtAl2026}. Dashed curves: C$^3$L with $\mc S_{1,3,5}$ and $N_{\rm pm}=6$.}
\label{fig:kondo-benchmarks}
\end{figure}

\emph{Additional tests in strongly correlated regimes at low temperature.}---C$^3$L is typically reliable at weak to intermediate correlation. To test its performance in strongly correlated regimes at low temperature, we consider the two-impurity Anderson model with interimpurity exchange
\begin{equation}
  \hat H_{\rm I} =- J^{\rm E} \hat{\bv S}_1\cdot  \hat{\bv S}_2 + \sum_{i=1,2}U\left(\hat n_{i\uparrow}-\frac12 \right)  \left(\hat n_{i\downarrow}-\frac12\right)
\end{equation}
at half filling, with $U=4$ and $\beta = 100$, where each impurity is Kondo-screened by an independent wide-band bath with a rectangular hybridization spectrum, following Ref.~\cite{LotemSonnerLinkEtAl2026}. We consider two cases: $J^{\rm E}=0$, where the system decouples into two single-impurity Anderson models, and $J^{\rm E} = 2$, where a fully screened effective impurity triplet gives rise to spin-$1$ Kondo behavior with a lower Kondo temperature. To resolve the low-temperature and  long-time impurity dynamics associated with the Kondo resonance, we fit the independent wide-band baths using a larger number of pseudomodes, $N_{\rm pm}  = 6$, and compute the spectral function in the $\mc S_{1,3,5}$ EOM space.

\cref{fig:kondo-benchmarks} shows that C$^3$L agrees well with the recently developed multiorbital tensor-network semigroup influence-matrix (SGIM) method of Ref.~\cite{LotemSonnerLinkEtAl2026} across the band, with small deviations in the Kondo-peak height. Resolving the narrow Kondo peak more precisely would benefit from preconditioned linear solvers at low frequencies (see Supplemental Material \cref{sec:EOM-CC}) and from more accurate low-frequency bath fits, potentially with more bath modes. Incorporating a more accurate treatment of static correlations, for example through multireference CC formulations, could extend C$^3$L to more strongly correlated regimes.

\clearpage
\thispagestyle{empty}
\onecolumngrid
\begin{bibunit}[apsrev4-2]
\makeatletter
\let\set@footnotewidth\set@footnotewidth@one
\let\compose@footnotes\compose@footnotes@one
\onecolumn@grid@setup
\def\@extra@b@citeb{SI}
\def\@extra@binfo{SI}
\@startbibunit
\immediate\write\@bibunitaux{\string\citation{REVTEX42Control,apsrev42Control}}
\makeatother
\begin{center}
\textbf{\large Supplemental Material for\\
Efficient simulations of spectra and transport for multiorbital quantum impurities
}
\\[\medskipamount]
Hao-En Li,$^{1,*}$ Zhen Huang,$^{2,3,\dagger}$ and Lin Lin$^{1,4,5,\ddagger}$\\
\smallskip
\small{\emph{$^1$\berkeleymath\\$^2$\ccq\\$^3$\ccm\\$^4$\caltechcms\\$^5$Applied Mathematics and Computational Research Division,\\ Lawrence Berkeley National Laboratory, Berkeley, CA 94720, USA}}\\
(Dated: October 28, 2026)

\end{center}
\begingroup
\makeatletter
\ltx@footnote@pop
\makeatother
\renewcommand{\thefootnote}{\fnsymbol{footnote}}
\footnotetext[1]{\href{mailto:haoen2021@math.berkeley.edu}{haoen2021@math.berkeley.edu}}
\footnotetext[2]{\href{mailto:zhuang@flatironinstitute.org}{zhuang@flatironinstitute.org}}
\footnotetext[3]{\href{mailto:lin@caltech.edu}{lin@caltech.edu}}
\endgroup
\setcounter{equation}{0}
\setcounter{figure}{0}
\setcounter{table}{0}
\setcounter{page}{1}
\setcounter{section}{0}
\setcounter{secnumdepth}{3}
\makeatletter
\renewcommand{\theequation}{S\arabic{equation}}
\renewcommand{\thefigure}{S\arabic{figure}}
\renewcommand{\thesection}{S\arabic{section}}
\renewcommand{\bibnumfmt}[1]{[S#1]}
\renewcommand{\citenumfont}[1]{S#1}

\onecolumngrid

\section{Hartree--Fock reference state and normal modes of the noninteracting Lindbladian}
\label{sec:HF}

The pseudomode HF calculation solves a self-consistent quasi-free Lindbladian in the full impurity-plus-pseudomode one-particle space. Starting from a 1-RDM $\gamma$, we construct $F[\gamma]$ in the impurity block, solve the quadratic Lyapunov equation for an updated 1-RDM, and iterate to self-consistency. The converged 1-RDM $\gamma_\star$ defines the Gaussian HF reference state $\rho_0$. Diagonalizing the corresponding noninteracting Lindbladian $\mc L_0$ then gives the normal modes $\{\bs P_\kappa,\widetilde{\bs P}_\kappa\}$.

A Lindbladian with a quadratic Hamiltonian, $H = \sum_{p,q} F_{pq} \hat c_p^\dag \hat c_q$, and a dissipator comprising linear fermionic jump operators (as in \cref{eq:dissipator}) generates quasi-free dynamics with a Gaussian steady state~\cite{BarthelZhang2022}. Its 1-RDM $\gamma_{qp} = \Tr(\hat c_p^\dagger \hat c_q \rho_{\rm ss})$ satisfies the Lyapunov equation
\begin{equation}\label{eq:lyapunov}
  -\I [F,\gamma] -\{\Gamma^{<} +\Gamma^{>}, \gamma\}  + 2\Gamma^{<} = 0.
\end{equation}
The Hartree--Fock approximation replaces the interaction by a quadratic mean-field Hamiltonian whose coefficients depend on the 1-RDM:
\begin{equation}
  F[\gamma]_{pq} = E_{pq} + \sum_{r,s} V_{pqrs} \gamma_{sr}.
\end{equation}
With this dependence, \cref{eq:lyapunov} becomes a nonlinear equation for $\gamma$. We iterate the map $\gamma\mapsto\gamma'$ defined by solving the Lyapunov equation at fixed $F[\gamma]$, using damping and Pulay mixing to accelerate convergence~\cite{Pulay1980,Pulay1982}. Spin-restricted HF calculations impose identical spin blocks and zero spin mixing at every iteration.

The Green's function of the quasi-free HF Lindbladian can be evaluated analytically. The Hamiltonian associated with the converged noninteracting Lindbladian $\mc L_0$ is
\begin{equation}
  H_{\rm CL,0} = \hat{\bv a} ^\dag \mathsf H_0 \hat{\bv a}: = \hat{\bv a}^\dag \mqty(F[\gamma_{\star}] & (\bv g^{<})^\dag & (\bv g^{>})^\dag\\ \bv g^{<} & H^{<}& 0 \\ \bv g^{>} & 0 & H^{>}) \hat{\bv a},\quad \hat{\bv a} := \left((\hat c_{i\sigma})_{i=1}^{n_{\rm I}+ N_{\rm pm}^< + N_{\rm pm}^> }
  \right)^{\top} := \left((\hat a_{i\sigma})_{i=1}^{n_{\rm I}},(\hat p_{i\sigma})_{i=1}^{N_{\rm pm}^<},(\hat d_{i\sigma})_{i=1}^{N_{\rm pm}^>}\right)^{\top}.
\end{equation}
For the dissipator, the corresponding block matrices are
 \begin{equation}
  \mathsf \Gamma^{<} = \mqty(0& 0& 0\\ 0 & \Gamma^{<} & 0\\ 0 & 0 & 0),\quad \mathsf \Gamma^{>} = \mqty(0& 0& 0\\ 0 & 0 & 0\\ 0 & 0 & \Gamma^{>}).
 \end{equation}
The odd-parity part $\underline{\mc L_0}$ of the $\ZZ_2$-graded generator is
 \begin{equation}
  \underline{\mc L_0}[\bullet] = -\I [H_{\rm CL,0},\bullet] -\sum_{i,j,\sigma} \Gamma^{<}_{ij} \left( 2\hat p_{i\sigma}^\dagger \bullet \hat p_{j\sigma} + \{\hat p_{j\sigma}\hat p_{i\sigma}^\dagger,\bullet\}\right) - \sum_{i,j,\sigma} \Gamma^{>}_{ij} \left( 2\hat d_{j\sigma} \bullet \hat d_{i\sigma}^\dagger + \{\hat d_{i\sigma}^\dagger\hat d_{j\sigma},\bullet\}\right).
 \end{equation}
For the noninteracting retarded Green's function $G^R_{0,pq}(t) = -\I \Theta(t) \Tr(\{\hat c_p(t),\hat c_q^\dagger(0)\}\rho_{0}) =: -\I \Theta(t) Y(t)$, we obtain
\begin{equation}
  \dv t Y(t) = \left[\Tr(\{ \underline{\mc L_0}^\dag[\hat c_p(t)],\hat c_q^\dagger(0)\}\rho_{0})\right]_{p,q} =  (-\I \mathsf H_0 - \mathsf \Gamma^{<} - \mathsf \Gamma^{>}) Y(t).
\end{equation}
Hence $G_0^R(t)=-\I\Theta(t)e^{(-\I\mathsf H_0-\mathsf\Gamma^<-\mathsf\Gamma^>)t}$. A Fourier transform gives
\begin{equation}
  \wt G_0^R(z) =  \int_0^\infty e^{\I z t} G_0^R(t) \dd t = (z - \mathsf H_0 + \I \mathsf \Gamma^{<} + \I \mathsf \Gamma^{>})^{-1} = \left[ z - \mqty(F[\gamma_\star]  & (\bv g^{<})^\dag & (\bv g^{>})^\dag\\ \bv g^{<} & H^{<} - \I \Gamma^{<} & 0 \\ \bv g^{>} & 0 & H^{>} - \I \Gamma^{>})\right]^{-1}.
\end{equation}
The impurity Green's function is the $(1,1)$ block of $\wt G_0^R(z)$. Taking the Schur complement gives
\begin{equation}
 \wt  G_{0,\rm I}^R(z) = \left(zI - F[\gamma_\star] - \Delta^{<}(z) - \Delta^{>}(z)\right)^{-1},\quad \Delta^{\gtrless}(z) = (\bv g^{\gtrless})^\dagger (zI - H^{\gtrless} + \I \Gamma^{\gtrless})^{-1} \bv g^{\gtrless}.
\end{equation}

We next construct the normal modes that define the HF reference for the post-HF CCSD calculation. The fermionic superoperators $\{\bs c_p,\bs c_p^\dagger,\wt{\bs c}_p,\wt{\bs c}_p^\dagger\}$ are defined by
\begin{equation}
        \bs c_p : \kett{\rho}   \mapsto \kett{\hat c_p \rho},\quad \bs c_p^\dagger : \kett{\rho} \mapsto \kett{\hat c_p^\dagger \rho},\quad
        \wt{\bs c}_p : \kett{\rho}  \mapsto  \kett{-\I \hat\Pi\rho \hat c_p^\dag \hat\Pi},\quad \wt{\bs c}_p^\dagger : \kett{\rho} \mapsto \kett{-\I \hat\Pi\rho \hat c_p \hat\Pi} .
\end{equation}
They satisfy the CAR because $\{\hat\Pi,\hat c_p\}=\{\hat\Pi,\hat c_p^\dagger\}=0$.

We partition the Lindbladian into noninteracting and interacting contributions,
\begin{equation}
    \bs K := \I \bs L = \bs K_0 + \bs W,
\end{equation}
where the noninteracting part $\bs K_0$ has the form $
    \bs K_0 = \sum_\sigma  \mathsf c_\sigma^\sharp \mathsf h_{\sigma}\mathsf c_\sigma $, with $\mathsf c_\sigma = (\bs c_{0\sigma},\cdots,\bs c_{n-1,\sigma},\wt{\bs c}_{0\sigma}^\dag ,\cdots,\wt{\bs c}_{n-1,\sigma}^\dag )^\top$, and $\mathsf c^\sharp_\sigma = (\bs c_{0\sigma}^\dagger,\cdots,\bs c_{n-1,\sigma}^\dagger,\wt{\bs c}_{0\sigma} ,\cdots,\wt{\bs c}_{n-1,\sigma} )$; $\mathsf h_\sigma$ is a $2n\times 2n$ matrix. Its component indices are $\zeta=0,\ldots,2n-1$.
 The coefficient matrix $\mathsf h_\sigma$
is generally nonnormal. Assuming it is diagonalizable, with $\mathsf h_\sigma V_\sigma=V_\sigma\mathrm{diag}(\varepsilon_{\xi,\sigma})_{\xi=0}^{2n-1}$, we define
    \begin{equation}
        \bs P^\star_{\xi\sigma} = \sum_{\zeta=0}^{2n-1} (V_\sigma^{-1})_{\xi\zeta} \mathsf c_{\zeta\sigma},\quad \widetilde{\bs P}^\star_{\xi\sigma} = \sum_{\zeta=0}^{2n-1} (V_\sigma)_{\zeta\xi} \mathsf c^\sharp_{\zeta\sigma}.
    \end{equation}
 These modes satisfy the biorthogonal CAR,  $\{\bs P^\star_{\xi\sigma}, \widetilde{\bs P}^\star_{\xi'\sigma'}\} = \delta_{\xi\xi'}\delta_{\sigma\sigma'}  $ and $\{\bs P^\star_{\xi\sigma},\bs P^\star_{\xi'\sigma'}\} = \{\widetilde{\bs P}^\star_{\xi\sigma},\widetilde{\bs P}^\star_{\xi'\sigma'}\} = 0$. We then perform the particle--hole transformation to define the normal modes $\bs P_{\xi\sigma}$ and $\widetilde{\bs P}_{\xi\sigma}$ and the corresponding eigenvalues $\lambda_{\xi,\sigma}$ and the particle--hole charge $q_{\xi,\sigma}$ as follows:
 \begin{equation}
    (\bs P_{\xi\sigma}, \widetilde{\bs P}_{\xi\sigma},\lambda_{\xi\sigma},q_{\xi\sigma}):= \begin{cases}
        (\bs P^\star_{\xi\sigma}, \widetilde{\bs P}^\star_{\xi\sigma},\varepsilon_{\xi\sigma},1), & \text{if } \Im \varepsilon_{\xi\sigma}<0,\\
        (\widetilde{\bs P}^\star_{\xi\sigma}, \bs P^\star_{\xi\sigma},-\varepsilon_{\xi\sigma},-1), & \text{if } \Im \varepsilon_{\xi\sigma}>0.
    \end{cases}
 \end{equation}
 The final modes $\{\bs P_{\xi\sigma}, \widetilde{\bs P}_{\xi\sigma}\}$ satisfy the same CAR as the $\bs P_{\xi\sigma}^\star$ and $\widetilde{\bs P}_{\xi\sigma}^\star$, and the Lindbladian superoperator $\bs K_0$ can be expressed in the diagonal form
  $
        \bs K_0 = \sum_{\kappa=1}^{4n}   \lambda_\kappa \widetilde{\bs P}_\kappa \bs P_\kappa,$ where $\kappa=(\sigma,\xi)\in\{\uparrow,\downarrow\}\times\{0,\ldots,2n-1\}$. In this representation, the Gaussian reference state $\kett{\rho_0}$ is the normal-mode vacuum:
        \begin{equation}\label{eq:normal_modes_vacuum}
            \bs P_\kappa\kett{\rho_0}=0, \quad \bbra{I} \widetilde{\bs P}_{\kappa } = 0,\quad \bbra{I}\kett{\rho_0} = \Tr(\rho_0)=1, \quad \forall \kappa.
        \end{equation}
The configurations $\kett{\Phi_S}$ are ordered products of normal-mode creators acting on the reference state:
        \begin{equation}\label{eq:config_right_S1}
          \kett{\Phi_S} = \boldsymbol \tau_S \kett{\rho_0} = \prod_{\kappa\in S}^{\nearrow} \widetilde{\bs P}_\kappa \kett{\rho_0},
        \end{equation}
        where $\nearrow$ denotes the ordering $(\uparrow,+)\prec (\uparrow,-)\prec (\downarrow,+)\prec (\downarrow,-)$ for the list $S = (\kappa_1 \prec \cdots\prec \kappa_{\abs{S}})$, with increasing $\xi$ within each sector and charge fixed by $q_{\xi\sigma}$, and $\boldsymbol \tau_S$ is the corresponding ordered product of the creation operators $\widetilde{\bs P}_\kappa$. Due to the $U(1)$-covariance of the Hamiltonian and each dissipator, the steady state $\rho_{\rm ss}$ satisfies $[\hat N_\sigma,\rho_{\rm ss}]=0$, which vectorizes to
       \begin{equation}
           (\bs N_\sigma - \wt{\bs N}_\sigma)\kett{\rho_{\rm ss}} = 0,\quad   \text{where } \bs N_\sigma = \sum_{i=0}^{n-1} \bs c_{i\sigma}^\dagger \bs c_{i\sigma},\quad  \wt{\bs N}_\sigma = \sum_{i=0}^{n-1} \wt{\bs c}_{i\sigma}^\dagger \wt{\bs c}_{i\sigma}.
       \end{equation}
 Thus the steady state lies in the charge-neutral sector of the doubled Fock space
  \begin{equation}
    \kett{\rho_{\rm ss}}\in \mathrm{Span}\{\kett{\Phi_S}: Q_\uparrow(S) = Q_\downarrow(S) = 0\},
  \end{equation}
  where the total particle--hole charge of a configuration $S$ is defined as $Q_{\uparrow/\downarrow}(S) = \sum_{\kappa\in S,\sigma(\kappa) = \uparrow/\downarrow} q_\kappa$.

The dual configuration is $\bbra{\wt\Phi_S}=\bbra{I}\boldsymbol\prod_{\kappa\in S}^{\searrow}\bs P_\kappa$, with the product ordered oppositely to that in \cref{eq:config_right_S1}. The left and right configurations are biorthogonal $\bbra{\wt\Phi_S}\kett{\Phi_{S'}}=\delta_{SS'}$.

\section{CCSD equations for interacting Lindbladians}
\label{sec:CCSD}
\subsection{Overview of the CCSD equations}
The exact steady state $\kett{\rho_{\rm ss}}$ of the interacting Lindbladian $\bs L = -\I \bs K$ satisfies
\begin{equation}
  \bs K \kett{\rho_{\rm ss}} = 0, \quad \bbra{I}\kett{\rho_{\rm ss}} = \Tr(\rho_{\rm ss}) = 1.
\end{equation}
CCSD uses the exponential ansatz
\begin{equation}
  \kett{\rho_{\rm CCSD}} = e^{\bs T }\kett{\rho_0},\quad \bs T = \bs T_1 +\bs T_2,\quad \bs T_r = \sum_{\abs{S} = 2r, Q_{\uparrow/\downarrow}(S) = 0} \mathrm{t}_S \boldsymbol \tau_S.
\end{equation}
Here $\bs T_1$ contains two normal-mode creators and $\bs T_2$ contains four. The subscripts $1,2$ count the particle--hole excitations: after the particle--hole transformation, the physical particle--hole excitation operator becomes the product of two normal-mode creators, with one carrying charge $q = +1$ and the other carrying $q=-1$. Specifically, let $a,b$ denote the $q=+1$ modes and $i,j$ denote the $q=-1$ modes:
\begin{equation}
  \bs T_1 = \sum_{\sigma, ai} \mathrm{t}_{ai}^\sigma \widetilde{\bs P}^+_{a\sigma} \widetilde{\bs P}^-_{i\sigma},\quad
  \bs T_2    =\sum_{aibj}\mathrm{t}^{\uparrow\downarrow}_{ai,bj}
 \widetilde{\bP}_{a\uparrow}^{+}\widetilde{\bP}_{i\uparrow}^{-}
 \widetilde{\bP}_{b\downarrow}^{+}\widetilde{\bP}_{j\downarrow}^{-}
 +\frac14 \sum_{\sigma abij}\mathrm{t}^{\sigma\sigma}_{ab,ij}
 \widetilde{\bP}_{a\sigma}^{+}\widetilde{\bP}_{b\sigma}^{+}
 \widetilde{\bP}_{i\sigma}^{-}\widetilde{\bP}_{j\sigma}^{-}.
\end{equation}
Interchanging two equal-sign creators changes the sign of their product, so the full amplitude tensors satisfy $\mathrm{t}_{ab,ij}^{\sigma\sigma} = -\mathrm{t}_{ba,ij}^{\sigma\sigma} = -\mathrm{t}_{ab,ji}^{\sigma\sigma}$. The number of independent complex amplitudes is
\begin{equation}
  \#{\bs T_1 \text{ amplitudes}} = 2 n^2,\quad \#{\bs T_2 \text{ amplitudes}} = n^4 + 2 \binom{n}{2}^2 ,\quad \#{\text{CCSD amplitudes}} = 2n^2 + n^4 + 2\binom{n}{2}^2.
\end{equation}
Every term in $\bs T$ contains at least one normal-mode creator. By \cref{eq:normal_modes_vacuum}, $\bbra{I}\bs T=0$ and $\bbra{I}e^{\bs T}=\bbra{I}$, so
\begin{equation}
  \Tr(\rho_{\rm CC}) = \bbra{I}\kett{\rho_{\rm CC}} = \bbra{I} e^{\bs T}\kett{\rho_0} = \bbra{I}\kett{\rho_0} = 1.
\end{equation}
The CC ansatz therefore preserves trace normalization without a separate constraint.

The CCSD amplitude equations follow by projecting the similarity-transformed residual $\kett{r(\bm{\mathrm{t}}_{\rm CCSD})}=\overline{\bs K}(\bm{\mathrm{t}}_{\rm CCSD})\kett{\rho_0}$ onto the excited test configurations, where $\overline{\bs K}(\bm{\mathrm{t}}_{\rm CCSD})=e^{-\bs T(\bm{\mathrm{t}}_{\rm CCSD})}\bs K e^{\bs T(\bm{\mathrm{t}}_{\rm CCSD})}$. Thus
\begin{equation}
  z_S(\bm{\mathrm{t}}_{\rm CCSD}) : =\bbra{\wt \Phi_S}\kett{r(\bm{\mathrm{t}}_{\rm CCSD})} =   \bbra{\wt \Phi_S} \overline{\bs K}(\bm{\mathrm{t}}_{\rm CCSD}) \kett{\rho_0} = 0,\quad \forall S\in \mc S_{\rm CCSD} := \{S: \abs{S} = 2,4, Q_{\uparrow/\downarrow}(S) = 0\}.
\end{equation}

\subsection{Derivation of the CCSD equations using exterior algebra}

The CCSD equations can be derived through several equivalent formalisms. One standard route is to normal-order the Baker--Campbell--Hausdorff (BCH) expansion in terms of particle--hole creation and annihilation operators and evaluate the resulting products via Wick's theorem~\cite{Wick1950,ShavitBartlett2009}. Alternatively, contractions can also be organized as Goldstone coupled-cluster diagrams~\cite{ShavitBartlett2009,BartlettMusial2007}. In the non-normal Lindbladian setting, however, both algebraic normal ordering and diagrammatic expansions demand additional bookkeeping \cite{CoveneyTew2026}. Because $\bP_\kappa$ is generally not the Hermitian adjoint of $\widetilde{\bP}_\kappa$, and the trace bra $\bbra{I}$ and Gaussian reference $\kett{\rho_0}$ constitute distinct left and right vacua, Wick contractions cannot exploit the standard Hermitian-adjoint symmetries of the Hamiltonian case. Similarly, a Goldstone diagrammatic formulation requires separate vertex tensors for distinct normal-ordered products. An exterior algebra formulation unifies these operations under a set of contraction rules. Here, we realize the CAR on an exterior tensor algebra where creation operators act via exterior multiplication and dual annihilation operators via interior contraction. Although uncommon in the traditional Hamiltonian CC literature, this approach connects directly to recent algebraic-geometric formulations of CC theory~\cite{GallierQuaintance2020,FaulstichOster2024,Sverrisdottir2026}.%

Wick's theorem remains valid for the superoperator CAR algebra, and the BCH expansion of $\overline{\bs K}$ terminates after four nested commutators, as in the Hamiltonian case~\cite[Sec. 10.1]{ShavitBartlett2009}:
\begin{equation}\label{eq:BCH_expansion}
  \overline{\bs K} = \bs K + [\bs K,\bs T] + \frac12 [[\bs K,\bs T],\bs T] + \frac16 [[[ \bs K,\bs T],\bs T],\bs T] + \frac1{24}[[[[\bs K,\bs T],\bs T],\bs T],\bs T].
\end{equation}
We first express the physical superoperators in terms of normal modes. For each component $\zeta$ of $\mathsf c_\sigma$ and $\mathsf c^\sharp_\sigma$, we sum over eigenmodes $\xi=0,\ldots,2n-1$ at fixed spin, with $\varepsilon_\xi=\varepsilon_{\xi\sigma}$:
\begin{equation}\label{eq:normal_modes_transform}
  \mathsf c_{\zeta\sigma} =\sum_{\Im \varepsilon_{\xi}<0} (V_\sigma)_{\zeta\xi} \bP_{\xi\sigma} + \sum_{\Im \varepsilon_{\xi}>0} (V_\sigma)_{\zeta\xi} \widetilde{\bP}_{\xi\sigma},\quad \mathsf c^\sharp_{\zeta\sigma} = \sum_{\Im \varepsilon_{\xi}<0} (V_\sigma^{-1})_{\xi\zeta} \widetilde{\bP}_{\xi\sigma} + \sum_{\Im \varepsilon_{\xi}>0} (V_\sigma^{-1})_{\xi\zeta} \bP_{\xi\sigma}.
\end{equation}
Each of $\{\bs c_p,\bs c_p^\dagger,\wt{\bs c}_p,\wt{\bs c}_p^\dagger\}$ is a linear combination of normal modes. Substituting \cref{eq:normal_modes_transform} into $\bs W=\bs W(\bs c_p,\bs c_p^\dagger,\wt{\bs c}_p,\wt{\bs c}_p^\dagger)$ gives a sum of factorized products,
\begin{equation}\label{eq:normal_modes_interacting}
  \bs W  =  \sum_{\alpha} g_\alpha \bs F_{\alpha1} \cdots \bs F_{\alpha\ell_\alpha},\quad \bs F_{\alpha j}  =\sum_{\kappa=1}^{4n} \left( v_{\alpha j,\kappa} \bP_\kappa + u_{\alpha j,\kappa} \widetilde{\bP}_\kappa\right),\quad v_{\alpha j,\kappa},u_{\alpha j,\kappa}\in \CC.
\end{equation}
Here $\ell_\alpha$ counts the factors in the $\alpha$th term of $\bs W$ and equals $2$ or $4$ for the present Lindbladian.

We introduce the exterior algebra on the doubled Fock space through the identities needed below; a general treatment is given in Ref.~\cite{GallierQuaintance2020}. Let $\mc V=\mc V_\uparrow\oplus\mc V_\downarrow=\CC^{4n}$, where $\mc V_\sigma=\CC^{2n}$ contains the normal modes of one spin. The indices $1,\ldots,4n$ enumerate both spin blocks, and each eigenmode fixes its charge $q_\kappa$. An antisymmetric tensor $a\in\wedge^r\mc V$ satisfies
\begin{equation}
  a_{i_{\pi(1)},\cdots,i_{\pi(r)}} = \mathrm{sgn}(\pi) a_{i_1,\cdots,i_r},\quad \forall \pi\in \mathfrak S_r, \quad\text{where}\quad i_1,\cdots,i_r\in \{1,\cdots,4n\}.
\end{equation}
Here $\mathfrak S_r$ is the symmetric group of $r$ elements and $\mathrm{sgn}(\pi)$ is the signature of the permutation $\pi$. Since the normal-mode operators satisfy the CAR, we can identify the $r$-tensor $a\in \bigwedge^r \mc V$ with the corresponding operator
\begin{equation}
  \bs A_r(a) = \frac1{r!} \sum_{i_1,\cdots,i_r=1}^{4n} a_{i_1,\cdots,i_r} \widetilde{\bs P}_{i_1}\cdots \widetilde{\bs P}_{i_r} = \sum_{\abs{S} = r } a_S \boldsymbol \tau_S.
\end{equation}
The subscript $r$ denotes the tensor degree. For $a\in\bigwedge^r\mc V$ and $b\in\bigwedge^q\mc V$, the wedge product is the $(r+q)$-tensor~\cite[Sec. 3.4]{GallierQuaintance2020}
\begin{equation}
  (a\wedge b)_{i_1,\cdots,i_{r+q}} :=  \sum_{\pi\in \mathfrak S_{r+q}, \pi(1)<\cdots<\pi(r), \pi(r+1)<\cdots<\pi(r+q)} \mathrm{sgn}(\pi) a_{i_{\pi(1)},\cdots,i_{\pi(r)}} b_{i_{\pi(r+1)},\cdots,i_{\pi(r+q)}}.
\end{equation}
This definition gives the algebra homomorphism
\begin{equation}
  \bs A_r(a) \bs A_q(b) = \bs A_{r+q}(a\wedge b),\quad \forall a\in {\bigwedge}^r \mc V, b\in {\bigwedge}^q \mc V.
\end{equation}
For a covector, i.e. linear functional $v=(v_1,\cdots,v_{4n})\in\mc V^\ast$, the interior contraction $\iota_v:\bigwedge^r\mc V\to\bigwedge^{r-1}\mc V$ is defined by~\cite[Sec. 3.6]{GallierQuaintance2020}
\begin{equation}
  (\iota_v a)_{i_1,\cdots,i_{r-1}} := \sum_{j=1}^{4n} v_j a_{j,i_1,\cdots,i_{r-1}},\quad \forall a\in{ \bigwedge}^r \mc V.
\end{equation}
The corresponding operator identities are
\begin{prop}\label{prop:interior_contraction}
For $v\in\mc V^\ast$ and $u\in\mc V$, define
\begin{equation}
  \widetilde{\bP}[u]:= \sum_{\kappa=1}^{4n} u_\kappa \widetilde{\bP}_\kappa,\quad \bP[v] := \sum_{\kappa=1}^{4n} v_\kappa \bs P_\kappa.
\end{equation}
Then
\begin{equation}
  \widetilde{\bP}[u] \bs A_r(a) = \bs A_{r+1}(u\wedge a),\quad \bP[v] \bs A_r(a) = \bs A_{r-1}(\iota_v a)  + (-1)^r \bs A_r(a) \bP[v].
\end{equation}
In particular,
\begin{equation}
  \bP[v] \bs A_r(a)\kett{\rho_0} = \bs A_{r-1}(\iota_v a)\kett{\rho_0}.
\end{equation}
\end{prop}
These identities follow directly from the definitions of the wedge product and interior contraction. They give the action of the similarity-transformed normal modes on $\kett{\rho_0}$:
\begin{prop}\label{prop:Fbar-action}
  Let $\bs F_{\alpha j} = \sum_{\kappa=1}^{4n} (v_{\alpha j,\kappa} \bP_\kappa + u_{\alpha j,\kappa} \widetilde{\bP}_\kappa) = \bP[v_{\alpha j}] + \widetilde{\bP}[u_{\alpha j}]$ (see \cref{eq:normal_modes_interacting}). Denote
  \begin{equation}\label{eq:w_definition}
    w_{\alpha j}^{(1)}= u_{\alpha j } + \iota_{v_{\alpha j}} \mathrm{t}^{(2)},\quad w_{\alpha j}^{(3)} = \iota_{v_{\alpha j}} \mathrm{t}^{(4)},
  \end{equation}
  where $\mathrm{t}^{(2)} \in \bigwedge^2 \mc V$ and $\mathrm{t}^{(4)} \in \bigwedge^4 \mc V$ are the antisymmetric tensors corresponding to the single and double amplitudes in CCSD. Then the action of the similarity-transformed operator $\overline{\bs F}_{\alpha j} = e^{-\bs T}\bs F_{\alpha j} e^{\bs T}$ on the reference state $\kett{\rho_0}$ is given by
  \begin{equation}
 \overline{\bs F}_{\alpha j}\bs A(a ) \kett{\rho_0} = \bs A(\overline{\mathfrak F}_{\alpha j}[a]) \kett{\rho_0},\quad  \text{where}\quad \overline{\mathfrak F}_{\alpha j}[a]
 =\iota_{v_{\alpha j}}a
 +(w_{\alpha j}^{(1)}+w_{\alpha j}^{(3)})\wedge a.
 \label{eq:Fbar-action}
  \end{equation}
Here $\bs A(a)=\sum_r\bs A_r(a^{(r)})$ for $a=\sum_r a^{(r)}$, with $a^{(r)}\in\bigwedge^r\mc V$.
\end{prop}
\begin{proof}

  We write the cluster operator as $\bs T = \bs A_2(\mathrm{t}^{(2)}) + \bs A_4(\mathrm{t}^{(4)})$. Since both components of $\bs T$ have even degrees, \cref{prop:interior_contraction} gives
  \begin{equation}
    [\bP[v_{\alpha j}],\bs T] = \bs A_1(\iota_{v_{\alpha j}}\mathrm{t}^{(2)}) + \bs A_3(\iota_{v_{\alpha j}}\mathrm{t}^{(4)}),\quad [\widetilde{\bP}[u_{\alpha j}],\bs T] = 0.
  \end{equation}
The first commutator $[\bs F_{\alpha j},\bs T]$ contains only odd-degree creator products, which commute with the even-degree creator products in $\bs T$. All higher nested commutators therefore vanish, and \cref{eq:BCH_expansion} gives
  \begin{equation}
    \overline{\bs F}_{\alpha j} = \bs F_{\alpha j} + [\bs F_{\alpha j},\bs T] = \bP[v_{\alpha j}] + \bs A(w_{\alpha j}^{(1)}+w_{\alpha j}^{(3)}).
  \end{equation}
Applying this identity to $\bs A(a)\kett{\rho_0}$ and using \cref{prop:interior_contraction} and the homomorphism property of $\bs A$ gives
  \begin{equation}
    \overline{\bs F}_{\alpha j}\bs A(a)\kett{\rho_0} = \bs A\left(\iota_{v_{\alpha j}}a+(w_{\alpha j}^{(1)}+w_{\alpha j}^{(3)})\wedge a\right)\kett{\rho_0},
  \end{equation}
  which proves \cref{eq:Fbar-action}.
\end{proof}
The quadratic superoperator $\bs K_0=\sum_{\kappa=1}^{4n}\lambda_\kappa\widetilde{\bP}_\kappa\bP_\kappa$ acts diagonally on $r$-tensors:
\begin{equation}\label{eq:K0_action}
  \bs K_0\bs A_r(a^{(r)})\kett{\rho_0} =: \bs A_r(\mathrm{D}_\lambda a^{(r)}) \kett{\rho_0},\quad (\mathrm{D}_\lambda a^{(r)})_{i_1,\cdots,i_r} =  \sum_{j=1}^r \lambda_{i_j} \cdot  a^{(r)}_{i_1,\cdots,i_r}.
\end{equation}
For any ordered interaction term $\bs F_{\alpha 1}\cdots \bs F_{\alpha \ell }$, define $\varrho_j$ (for $j=1,\cdots,\ell,\ell+1$) by
\begin{equation}\label{eq:varrho_definition}
  \overline{\bs F}_{\alpha j }\cdots \overline{\bs F}_{\alpha \ell}\kett{\rho_0} = \bs A(\varrho_{\alpha, j})\kett{\rho_0},\quad j=1,\cdots,\ell,\quad \varrho_{\alpha, \ell+1}=1.
\end{equation}
By \cref{prop:Fbar-action}, these tensors satisfy the recursion
\begin{equation}\label{eq:varrho_recursion}
  \varrho_j^{(r)} = \iota_{v_j} \varrho_{j+1}^{(r+1)} + w_j^{(1)}\wedge \varrho_{j+1}^{(r-1)} + w_j^{(3)}\wedge \varrho_{j+1}^{(r-3)},\quad j=1,\cdots,\ell.
\end{equation}
We suppress $\alpha$ in \cref{eq:varrho_recursion} and the next few equations.
Components of negative degree vanish. For a quadratic term $\bs F_{\alpha1}\bs F_{\alpha2}$, $\ell=2$ and \cref{eq:varrho_recursion} gives
\begin{equation}
  \varrho_2^{(1)}=w_2^{(1)},\quad  \varrho_2^{(3)}=w_2^{(3)},\quad
  \varrho_1^{(2)}=\iota_{v_1}w_2^{(3)}+w_1^{(1)}\wedge w_2^{(1)},\quad
  \varrho_1^{(4)}=w_1^{(1)}\wedge w_2^{(3)}
  +w_1^{(3)}\wedge w_2^{(1)}.
\end{equation}
For a quartic interaction term $\bs F_{\alpha 1}\bs F_{\alpha 2}\bs F_{\alpha 3}\bs F_{\alpha 4}$ ($\ell = 4$), we have
\begin{equation}
 \begin{aligned}
  &\varrho_4^{(1)}=w_4^{(1)},\quad
  \varrho_4^{(3)}=w_4^{(3)},\quad
  \varrho_3^{(0)}=\iota_{v_3}\varrho_4^{(1)},\quad
  \varrho_3^{(2)}=\iota_{v_3}\varrho_4^{(3)}+w_3^{(1)}\wedge\varrho_4^{(1)},\quad
  \varrho_3^{(4)}=w_3^{(1)}\wedge\varrho_4^{(3)}+w_3^{(3)}\wedge\varrho_4^{(1)},\\[-2pt]
  &\varrho_3^{(6)}=w_3^{(3)}\wedge\varrho_4^{(3)},\quad
  \varrho_2^{(1)}=\iota_{v_2}\varrho_3^{(2)}+w_2^{(1)}\wedge\varrho_3^{(0)},\quad
  \varrho_2^{(3)}=\iota_{v_2}\varrho_3^{(4)}+w_2^{(1)}\wedge\varrho_3^{(2)}+w_2^{(3)}\wedge\varrho_3^{(0)},\\[-2pt]
  &\varrho_2^{(5)}=\iota_{v_2}\varrho_3^{(6)}+w_2^{(1)}\wedge\varrho_3^{(4)}+w_2^{(3)}\wedge\varrho_3^{(2)},\quad
  \varrho_1^{(2)}=\iota_{v_1}\varrho_2^{(3)}+w_1^{(1)}\wedge\varrho_2^{(1)},\quad  \varrho_1^{(4)}=\iota_{v_1}\varrho_2^{(5)}+w_1^{(1)}\wedge\varrho_2^{(3)}+w_1^{(3)}\wedge\varrho_2^{(1)}.
 \end{aligned}
 \label{eq:quartic}
\end{equation}
Only $\varrho_1^{(2)}$ and $\varrho_1^{(4)}$ enter the projected CCSD residual. Combining \cref{eq:varrho_definition,eq:varrho_recursion,eq:normal_modes_interacting} with the diagonal action in \cref{eq:K0_action} gives
\begin{equation}
  \kett{r(\bm{\mathrm{t}}_{\rm CCSD})} = \bs A\left(r \right)\kett{\rho_0},\quad r^{(2)} = \mathrm{D}_\lambda \mathrm{t}^{(2)}   + \sum_\alpha g_\alpha \varrho_{\alpha, 1}^{(2)}  ,\quad r^{(4)} = \mathrm{D}_\lambda \mathrm{t}^{(4)} + \sum_\alpha g_\alpha  \varrho_{\alpha, 1}^{(4)} .
\end{equation}
Projection onto $\{\bbra{\wt\Phi_S}:S\in\mc S_{\rm CCSD}\}$ selects the charge-neutral degree-$2$ and degree-$4$ components. The CCSD equations are therefore
\begin{equation}\label{eq:CCSD_exterior_algebra}
  \eval{\left(\mathrm{D}_\lambda \mathrm{t}^{(2)} + \sum_\alpha g_\alpha  \varrho_{\alpha, 1}^{(2) } \right)}_{Q = 0 }  = 0,\quad \eval{\left( \mathrm{D}_\lambda \mathrm{t}^{(4)} + \sum_\alpha g_\alpha  \varrho_{\alpha, 1}^{(4)} \right) }_{Q = 0 }  = 0.
\end{equation}
This gives the exterior-algebra formulation of the CCSD equations.

Several optimizations could be applied to reduce the cost of these recursions. For instance,
only contraction paths that can reach a charge-neutral degree-$2$ or degree-$4$ tensor need to be evaluated. An intermediate can carry nonzero charge, but it can be discarded if the remaining factors cannot return it to $Q=0$. Contracting before taking exterior products also avoids high-rank intermediates. For example, for $3$-tensors $a$ and $b$, direct evaluation of $\iota_{v_1}\iota_{v_2}(a\wedge b)$ forms tensors of degrees six and five. Instead, the Leibniz rule
gives~\cite[Sec. 3.6]{GallierQuaintance2020}
\begin{equation}
\iota_{v_1}\iota_{v_2}(a\wedge b)
= (\iota_{v_1}\iota_{v_2}a)\wedge b
+(\iota_{v_2}a)\wedge(\iota_{v_1}b) -(\iota_{v_1}a)\wedge(\iota_{v_2}b)
+a\wedge(\iota_{v_1}\iota_{v_2}b),
\end{equation}
so only tensors of degree at most four need to be
formed. Moreover, paths that share an ordered prefix generate identical intermediate tensors, which need to be evaluated only once and can be reused across multiple paths. The optimized contractions
can be represented as array operations and compiled into GPU kernels
for efficient evaluation.
The same contraction graph provides the Jacobian--vector products required by the Newton--Krylov solver described in \cref{sec:CCSD_solver}.

\subsection{Solving the CCSD equations with Newton--Krylov iterations}\label{sec:CCSD_solver}

We solve the nonlinear system \cref{eq:CCSD_exterior_algebra} with a Newton--Krylov method, which has been shown to improve the convergence behavior of Hamiltonian CC equations~\cite{YangBrabecVeisEtAl2020}. By adapting the forcing term at each Newton step, the Eisenstat--Walker strategy~\cite{EisenstatWalker1996} dynamically sets the Krylov tolerance to prevent oversolving. The required Jacobian--vector product (JVP) acts on the independent charge-neutral components of the degree-$2$ and degree-$4$ amplitude tensors. In practice, the implementation packs these complex amplitudes into a real vector. The derivation below only uses complex notation for clarity.

Differentiating \cref{eq:w_definition} with respect to the CCSD amplitudes gives
 \begin{equation}
  \delta w_{\alpha j}^{(1)} = \iota_{v_{\alpha j}}\delta \mathrm{t}^{(2)},\quad \delta w_{\alpha j}^{(3)} = \iota_{v_{\alpha j}}\delta \mathrm{t}^{(4)}.
 \end{equation}
Applying the product rule to \cref{eq:varrho_recursion} gives
\begin{equation}
  \delta \varrho_j^{(r)} = \iota_{v_j} \delta \varrho_{j+1}^{(r+1)} + \delta w_j^{(1)}\wedge \varrho_{j+1}^{(r-1)} + w_j^{(1)}\wedge \delta \varrho_{j+1}^{(r-1)} + \delta w_j^{(3)}\wedge \varrho_{j+1}^{(r-3)} + w_j^{(3)}\wedge \delta \varrho_{j+1}^{(r-3)},\quad j=1,\cdots,\ell,\quad \delta \varrho_{\ell+1}^{(0)} = 0.
\end{equation}
The resulting JVP is
\begin{equation}
   \mathscr{D}
    \bv z(\bm{\mathrm{t}}_{\rm CCSD})(\delta \mathrm{t}) = \eval{\left(\mathrm{D}_\lambda \delta\mathrm{t} + \sum_\alpha g_\alpha  \delta \varrho_{\alpha,1}\right)}_{Q = 0, \abs{S} = 2,4} ,\quad \bv z(\bm{\mathrm{t}}_{\rm CCSD}) =  [z_S(\bm{\mathrm{t}}_{\rm CCSD})]_{S\in \mc S_{\rm CCSD}}.
\end{equation}
The Newton vector contains the same independent charge-neutral degree-$2$ and degree-$4$ components as $\bm{\mathrm{t}}_{\rm CCSD}$. At each Newton step, GMRES solves
\begin{equation}
  \mathscr{D} \bv z[\bm{\mathrm{t}}_{\rm CCSD}^{ k }](\delta \bm{\mathrm{t}}^{ k }) = -\bv z(\bm{\mathrm{t}}_{\rm CCSD}^{ k  }).
\end{equation}
We left-precondition Newton--GMRES with $\mathrm{D}_\lambda^{-1}$, using the quadratic-reference diagonal $(\mathrm{D}_\lambda)_{SS}=\sum_{\kappa\in S}\lambda_\kappa$. The stopping criterion below uses the unpreconditioned residual.
We terminate GMRES and update the amplitudes according to~\cite{DemboEisenstatSteihaug1982,EisenstatWalker1994,EisenstatWalker1996}
\begin{equation}
  \left\|\mathscr{D} \bv z[\bm{\mathrm{t}}_{\rm CCSD}^{k}](\delta \bm{\mathrm{t}}^{k})+\bv z(\bm{\mathrm{t}}_{\rm CCSD}^{k})\right\| \leq \eta_k\left\|\bv z(\bm{\mathrm{t}}_{\rm CCSD}^{k})\right\|,\quad 0\leq \eta_k<1,\quad \bm{\mathrm{t}}_{\rm CCSD}^{k+1}=\bm{\mathrm{t}}_{\rm CCSD}^{k}+ s_k \delta
  \bm{\mathrm{t}}^{k}.
\end{equation}
The adaptive forcing term allows a loose inner tolerance when the nonlinear residual is large and tightens the GMRES solve near convergence. This avoids oversolving the early Newton steps while preserving rapid local convergence. A backtracking line search with an Armijo condition determines $s_k$ and enforces sufficient residual decrease~\cite{Armijo1966}.

\section{EOM-CC equations for Green's functions}\label{sec:EOM-CC}

We derive the EOM-CC equations for the Lindbladian Green's function~\cite{StantonBartlett1993,ShavitBartlett2009}. For impurity sites $i,j\in\mc I$, the retarded Green's function is
\begin{equation}
  G_{ij,\sigma}^R(t) = -\I \Theta(t) \Tr \left[ \rho_{\rm CC} \{\hat c_{i\sigma}(t), \hat c_{j\sigma}^\dag (0)\}\right],\quad \hat c_{i\sigma}(t) = e^{t\underline{\mc L}^\dag } [\hat c_{i\sigma}]
\end{equation}
where the odd-parity $\ZZ_2$-graded generator $\underline{\mc L}[\bullet]:=\hat\Pi\mc L[\hat\Pi\cdot\bullet]$ accounts for the fermionic signs associated with the environment~\cite{SchwarzGoldsteinDordaEtAl2016,CirioKuoChenEtAl2022,HuangZhuParkLin2025,BertrandBesserveFerreroEtAl2025}. Its explicit form is
\begin{equation}
  \underline{\mc L} [\bullet ] =  - \I[\hat H_{\rm CL},\bullet]  - \sum_{i,j,\sigma} \Gamma^{<}_{ij} \left(2\hat p_{i\sigma}^\dagger \bullet \hat p_{j\sigma} + \{\hat p_{j\sigma}\hat p_{i\sigma}^\dagger,\bullet\}\right) -\sum_{i,j,\sigma} \Gamma^{>}_{ij} \left(2\hat d_{j\sigma} \bullet \hat d_{i\sigma}^\dag + \{\hat d_{i\sigma}^\dag \hat d_{j\sigma},\bullet\}\right).
\end{equation}
Therefore,
\begin{equation}
  \begin{aligned}
  G_{ij,\sigma}^R(t) &= -\I \Theta(t) \Tr \left[ \rho_{\rm CC} \{e^{t\underline{\mc L}^\dag } [\hat c_{i\sigma}], \hat c_{j\sigma}^\dag  \}\right]  = -\I \Theta(t) \Tr \left[  \hat c_{i\sigma} e^{t\underline{\mc L}}(\hat c_{j\sigma}^\dag \rho_{\rm CC}  +  \rho_{\rm CC} \hat c_{j\sigma}^\dag ) \right]\\
   &  =   -\I \Theta(t) \bbra{I}\boldsymbol c_{i\sigma} e^{ {-\I t\boldsymbol K}} \mathbf P_{\rm o}  (\bs c_{j\sigma}^\dag -\I \widetilde{\bs c}_{j\sigma} )\kett{\rho_{\rm CC}} = -\I \Theta(t) \bbra{I}\boldsymbol c_{i\sigma} e^{ {-\I t\boldsymbol K}} \mathbf P_{\rm o}  (\bs c_{j\sigma}^\dag -\I \widetilde{\bs c}_{j\sigma} )e^{\bs T [\bv{ {t}} _{\rm CC }]}\kett{\rho_0}\\
   & = -\I \Theta(t) \bbra{I}\boldsymbol c_{i\sigma} e^{\boldsymbol{T}[\bv{ {t}}_{\rm CC}]} e^{ {-\I t\overline{\boldsymbol K}}} \mathbf P_{\rm o}  (\overline{\bs c_{j\sigma}^\dagger} -\I \overline{\widetilde{\bs c}_{j\sigma}} ) \kett{\rho_0}.
  \end{aligned}
\end{equation}
Here $\bv P_{\rm o}=\frac12(\bs I-(-1)^{\bs N+\wt{\bs N}})$ projects onto the odd-parity subspace of the doubled Fock space. $\underline{\mathcal L}^\dag$ denotes the Hilbert--Schmidt adjoint of $\underline{\mathcal L}$. The even-parity cluster operator $\boldsymbol T$ commutes with $\mathbf P_{\rm o}$.

EOM-CC restricts the response to an odd-parity subspace with a specified excitation-rank cutoff. For the retarded Green's function, we use the charge-$1$ sector with maximum rank $3$ or $5$:
\begin{equation}
  \mc S_{\mc R}^{(\sigma)} = \{S:\abs{S} \in \mc R, Q_{\sigma}(S) = 1, Q_{\overline{\sigma}}(S) = 0\},\quad \mc R = \{1,3\}\text{ or }\{1,3,5\}.
\end{equation}
Let $\mathbf P_{\mc S_{\mc R}^{(\sigma)}}$ be the corresponding oblique projection superoperator onto the subspace spanned by $\{\kett{\Phi_S}: S\in \mc S_{\mc R}^{(\sigma)}\}$.
\begin{equation}
  \mathbf P_{\mc S_{\mc R}^{(\sigma)}} = \sum_{S\in \mc S_{\mc R}^{(\sigma)}} \kett{\Phi_S}\bbra{\wt \Phi_S}.
\end{equation}
The EOM amplitudes $x_{j\sigma}=\sum_{r\in\mc R}x_{j\sigma}^{(r)}$ are represented by antisymmetric tensors $x_{j\sigma}^{(r)}\in\bigwedge^r\mc V$. We write
 $
    \boldsymbol{A}(x_{j\sigma}) \kett{\rho_0} = \sum_{r\in \mc R} \bs A_r(x_{j\sigma}^{(r)})\kett{\rho_0}.
$ We approximate the time-evolved state $e^{-\I t\overline{\boldsymbol K}}\mathbf P_{\rm o}(\overline{\bs c_{j\sigma}^\dagger}-\I\overline{\widetilde{\bs c}_{j\sigma}})\kett{\rho_0}$ by $\boldsymbol A(x_{j\sigma}(t))\kett{\rho_0}$ and impose the evolution equation in the test EOM space $\{\bbra{\wt\Phi_S}:S\in\mc S_{\mc R}^{(\sigma)}\}$:
  \begin{equation}
    \mathbf P_{\mc S_{\mc R}^{(\sigma)}}\left[\I\boldsymbol A(\partial_t x_{j\sigma}(t))-\overline{\boldsymbol K}\boldsymbol A(x_{j\sigma}(t))\right]\kett{\rho_0}=0.
  \end{equation}
  Since $\boldsymbol A(x_{j\sigma}(t))\kett{\rho_0}$ lies in the retained space, this gives
  \begin{equation}
    \boldsymbol{A}(\partial_t x_{j\sigma}(t))\kett{\rho_0} = -\I \mathbf P_{\mc S_{\mc R}^{(\sigma)}} \overline{\boldsymbol K} \boldsymbol{A}(x_{j\sigma}(t))\kett{\rho_0} = -\I \mathbf P_{\mc S_{\mc R}^{(\sigma)}} [\overline{\boldsymbol K}, \boldsymbol{A}(x_{j\sigma}(t))]\kett{\rho_0} -\I \mathbf P_{\mc S_{\mc R}^{(\sigma)}} \boldsymbol{A}(x_{j\sigma}(t)) \overline{\boldsymbol K}\kett{\rho_0}.
  \end{equation}
The last term contains the static residual $\overline{\boldsymbol K}\kett{\rho_0}$. Its scalar component vanishes by trace preservation, and its rank-$2$ and rank-$4$ components vanish at a converged CCSD solution. The remaining components have rank at least six. Multiplication by the pure-creator polynomial $\boldsymbol A(x)$ could only raise the rank, so this term has zero projection onto either $\mc S_{1,3}$ or $\mc S_{1,3,5}$. The resulting initial-value problem is
  \begin{equation}\label{eq:EOM_equation}
    \boldsymbol{A}(\partial_t x_{j\sigma}(t))\kett{\rho_0} = -\I \mathbf P_{\mc S_{\mc R}^{(\sigma)}} [\overline{\boldsymbol K}, \boldsymbol{A}(x_{j\sigma}(t))]\kett{\rho_0},\quad  \boldsymbol{A}(x_{j\sigma}(0))\kett{\rho_0} = \mathbf P_{\mc S_{\mc R}^{(\sigma)}}    (\overline{\bs c_{j\sigma}^\dagger} -\I \overline{\widetilde{\bs c}_{j\sigma}} ) \kett{\rho_0}.
  \end{equation}
Define the linear map $\mathrm A_{\mc R,\sigma}$ and initial vector $b_{j\sigma}^{\mc R}$ by
  \begin{equation}
   [ \mathrm A_{\mc R,\sigma} x]_S := \bbra{\wt \Phi_S} [\overline{\boldsymbol K}, \boldsymbol{A}(x)]\kett{\rho_0},\quad [b^{\mc R}_{j\sigma}]_S := \bbra{\wt \Phi_S}  (\overline{\bs c_{j\sigma}^\dagger} -\I \overline{\widetilde{\bs c}_{j\sigma}} ) \kett{\rho_0},\quad S\in \mc S_{\mc R}^{(\sigma)}.
  \end{equation}
Projecting \cref{eq:EOM_equation} onto $\bbra{\wt\Phi_S}$ yields
  \begin{equation}
    \partial_t x_{j\sigma}(t) = -\I \mathrm A_{\mc R,\sigma}(x_{j\sigma}(t)),\quad x_{j\sigma}(0) = b_{j\sigma}^{\mc R},\quad \text{thus} \quad x_{j\sigma}(t) = e^{-\I t \mathrm A_{\mc R,\sigma} } b_{j\sigma}^{\mc R}.
  \end{equation}
Reconstruction of the Green's function also requires the row vectors
  \begin{equation}
  [s_{i \sigma}^{\mc R}]_S := \bbra{I}\boldsymbol c_{i\sigma} \kett{\Phi_S},\quad S\in \mc S_{\mc R}^{(\sigma)},\quad i\in \mc I.
  \end{equation}
The EOM-CC retarded Green's function is then
  \begin{equation}\label{eq:EOM-Green-function}
    \begin{aligned}
    G_{ij,\sigma}^{R,\mc R}(t) &:= -\I \Theta(t)\bbra{I}\boldsymbol c_{i\sigma}e^{\bs T}\boldsymbol A(x_{j\sigma}(t))\kett{\rho_0} = -\I\Theta(t)\sum_{S\in\mc S_{\mc R}^{(\sigma)}}\bbra{I}\boldsymbol c_{i\sigma}e^{\bs T}\kett{\Phi_S}\,[x_{j\sigma}(t)]_S\\
    & = -\I\Theta(t)\sum_{S\in\mc S_{\mc R}^{(\sigma)}}\bbra{I}\boldsymbol c_{i\sigma}\kett{\Phi_S}\,[x_{j\sigma}(t)]_S
    = -\I\Theta(t)\left(s_{i\sigma}^{\mc R}\right)^\top e^{-\I t\mathrm A_{\mc R,\sigma}}b_{j\sigma}^{\mc R}.
    \end{aligned}
  \end{equation}
  To obtain the penultimate equality, we apply \cref{prop:Fbar-action} to the linear superoperator $\bs c_{i\sigma}$, giving $\overline{\bs c_{i\sigma}}=\bs c_{i\sigma}+[\bs c_{i\sigma},\bs T]$. The commutator contains only products of creators and is therefore annihilated by $\bbra{I}$. Combining this with $\bbra{I}e^{\bs T}=\bbra{I}$ yields
  \begin{equation}
    \bbra{I}\bs c_{i\sigma}e^{\bs T}=\bbra{I}\overline{\bs c_{i\sigma}}=\bbra{I}\bs c_{i\sigma}.
  \end{equation}
A Fourier transform of \cref{eq:EOM-Green-function} gives the frequency-domain EOM-CC Green's function
  \begin{equation}\label{eq:EOM-Green-function-frequency}
    \wt G_{ij,\sigma}^{R,\mc R}(z) = \int_{-\infty}^{+\infty} G_{ij,\sigma}^{R,\mc R}(t) e^{\I z t} \dd t = \left(s_{i\sigma}^{\mc R}\right)^\top (zI-\mathrm A_{\mc R,\sigma})^{-1}b_{j\sigma}^{\mc R},\quad \Im z>\max\Im\operatorname{spec}(\mathrm A_{\mc R,\sigma}).
  \end{equation}
Real-frequency values are evaluated from the resolvent with the stated broadening. We evaluate $\mathrm A_{\mc R,\sigma}x_{j\sigma}$ by tensor contractions without forming $\mathrm A_{\mc R,\sigma}$ explicitly, and solve the resulting linear systems with Krylov methods. To derive the contraction formula, consider an interaction term $\bs F_{\alpha1}\cdots\bs F_{\alpha\ell}$ and define $\varsigma_j[x]$ for $j=1,\cdots,\ell+1$ by
  \begin{equation}
    \overline{\bs F}_{\alpha j }\cdots \overline{\bs F}_{\alpha \ell} \boldsymbol{A}(x) \kett{\rho_0} = \boldsymbol{A}(\varsigma_{\alpha, j}[x])\kett{\rho_0},\quad j=1,\cdots,\ell,\quad \varsigma_{\alpha, \ell+1}[x] =x.
  \end{equation}
The corresponding recursion is
  \begin{equation}
    \varsigma_j^{(r)}[x] = \iota_{v_j} \varsigma_{j+1}^{(r+1)}[x] + w_j^{(1)}\wedge \varsigma_{j+1}^{(r-1)}[x] + w_j^{(3)}\wedge \varsigma_{j+1}^{(r-3)}[x],\quad j=1,\cdots,\ell
  \end{equation}
This recursion has the form of \cref{eq:varrho_recursion}, but is initialized with the odd EOM tensor $\varsigma_{\ell+1}[x]=x$. The commutator acting on $\kett{\rho_0}$ is therefore
  \begin{equation}
    [\overline{\bs F}_{\alpha 1}\cdots \overline{\bs F}_{\alpha \ell}, \boldsymbol{A}(x)]\kett{\rho_0} = \boldsymbol{A}(\varsigma_{\alpha, 1}[x])\kett{\rho_0} - \boldsymbol{A}(x)\overline{\bs F}_{\alpha 1}\cdots \overline{\bs F}_{\alpha \ell}\kett{\rho_0} = \boldsymbol{A}(\varsigma_{\alpha, 1}[x] - x \wedge \varrho_{\alpha, 1})\kett{\rho_0}.
  \end{equation}
  Therefore, for $\bs K = \sum_{\kappa=1}^{4n} \lambda_\kappa\widetilde{\bP}_\kappa \bP_\kappa + \sum_\alpha g_\alpha \bs F_{\alpha 1}\cdots \bs F_{\alpha \ell}$, we have
  \begin{equation}
    [\overline{\bs K}_0, \bs A(x)] \kett{\rho_0} = [\bs K_0 + \bs A(\mathrm{D}_\lambda \mathrm t), \bs A(x)]\kett{\rho_0} = \bs A(\mathrm{D}_\lambda x + (\mathrm{D}_\lambda \mathrm t) \wedge x- x \wedge  (\mathrm D_\lambda \mathrm t))\kett{\rho_0} = \bs A(\mathrm{D}_\lambda x  )\kett{\rho_0}.
  \end{equation}
Since $x$ has odd degree and $\mathrm D_\lambda\mathrm t$ has even degree, $x\wedge(\mathrm D_\lambda\mathrm t)=(\mathrm D_\lambda\mathrm t)\wedge x$~\cite[Sec. 3.4]{GallierQuaintance2020}. The projected commutator can therefore be evaluated as
\begin{equation}\label{eq:action_eom}
 \bv P_{\mc S_{\mc R}^{(\sigma)}} [\overline{\bs K}, \boldsymbol{A}(x)]\kett{\rho_0} =\bs A
 \left[\eval{\left( \mathrm{D}_\lambda x + \sum_{\alpha}g_\alpha\big( \varsigma_{\alpha, 1}[x] - x \wedge \varrho_{\alpha, 1} \big)\right)}_{S \in \mc S_{\mc R}^{(\sigma)}}\right]\kett{\rho_0}.
\end{equation}
In coefficient space, \cref{eq:action_eom} gives the linear EOM map
\begin{equation}
(\mathrm A_{\mc R,\sigma} x)_{S\in \mc S_{\mc R}^{(\sigma)}} = \eval{\left( \mathrm{D}_\lambda x + \sum_{\alpha}g_\alpha\big( \varsigma_{\alpha, 1}[x] - x \wedge \varrho_{\alpha, 1} \big)\right)}_{S \in \mc S_{\mc R}^{(\sigma)}}.
\end{equation}
These contractions can be optimized using the same strategies as for the CCSD residual.

The similarity transformation of the
linear odd operator $\bs c_{j\sigma}^\dagger-\I\widetilde{\bs c}_{j\sigma}$ by the rank-$2$/rank-$4$ cluster operator $\bs T$ terminates after the first commutator, so its action on $\kett{\rho_0}$ produces only rank-$1$ and rank-$3$ components. The linear sink $\bbra{I}\bs c_{i\sigma}$ has support only at rank $1$. Thus, neither $b_{j\sigma}^{\mc R}$ nor $s_{i\sigma}^{\mc R}$ has a rank-$5$ component:
 \begin{equation}
  b^{1,3,5}_{j\sigma} = \mqty(b^{1,3}_{j\sigma} \\ 0),\quad (s^{1,3,5}_{i\sigma})^\top = \mqty((s^{1,3}_{i\sigma})^\top & 0).
\end{equation}To solve $(zI-\mathrm A_{\mc R,\sigma})x_{j\sigma}=b_{j\sigma}^{\mc R}$ with GMRES, we construct the $m_{\rm K}$-dimensional Krylov space $\mc K_{m_{\rm K}}(zI-\mathrm A_{\mc R,\sigma},b_{j\sigma}^{\mc R})$ by Arnoldi iteration with Gram--Schmidt orthogonalization. Shift invariance gives $\mc K_{m_{\rm K}}(zI-\mathrm A_{\mc R,\sigma},b_{j\sigma}^{\mc R})=\mc K_{m_{\rm K}}(\mathrm A_{\mc R,\sigma},b_{j\sigma}^{\mc R})$ for any $z\in\mathbb C$. The basis therefore needs to be constructed only
once for each source $j\sigma$ and can be reused
across frequencies $z$ and output indices $i\sigma$ when evaluating $G_{ij,\sigma}^{R,\mc R}(z)$.
The shared, shift-invariant Arnoldi pass is unpreconditioned. For the strongly correlated two-impurity Anderson benchmark in the End Matter, separate frequency-specific GMRES refinements use right preconditioning by $(zI-\mathrm{D}_\lambda)^{-1}$ within the EOM space.
\section{Numerical setups and additional results}\label{sec:numerical_setups}
Here we specify the models and numerical settings for the calculations in the main text and present additional results. Unless otherwise stated, bath fits were obtained with the \texttt{RealTimeBath} package~\cite{RealTimeBath}, using the semidefinite-programming (SDP) construction of Ref.~\cite{HuangParkChanLin2026}. For fermionic baths, the fitted lesser and greater hybridization functions $\Delta^{\gtrless}(t)$ are represented by coupled-Lindblad pseudomodes.
The target functions are obtained from the positive-semidefinite, matrix-valued spectral density $J(\omega) = \left[J_{ij}(\omega)\right]_{i,j=1}^{n_{\rm I}}$ via
\begin{equation}\label{eq:gtrless}
\Delta^{\gtrless}(t) = \int_{-\infty}^{+\infty} J(\omega) f^{\gtrless}(\omega) e^{-\I \omega t} \dd \omega .
\end{equation}
Here, $f^<(\omega) = f_{\rm FD}^{\beta,\mu}(\omega)$ and $f^>(\omega) = 1 - f_{\rm FD}^{\beta,\mu}(\omega)$, where $
f_{\rm FD}^{\beta,\mu}(\omega) = \frac{1}{e^{\beta (\omega - \mu)} + 1}
 $
is the Fermi--Dirac distribution for a reservoir at inverse temperature $\beta$ and chemical potential $\mu$. For nonequilibrium systems with multiple independently thermalized reservoirs $r$, the global-equilibrium relation in \cref{eq:gtrless} is generalized by replacing $J(\omega) f^{\gtrless}(\omega)$ with $\sum_r J_r(\omega) f^{\gtrless}_r(\omega)$, using the corresponding $f_{\rm FD}^{\beta_r,\mu_r}(\omega)$ for each reservoir.

\begin{figure}[!h]
\centering
\includegraphics[width=0.6\linewidth]{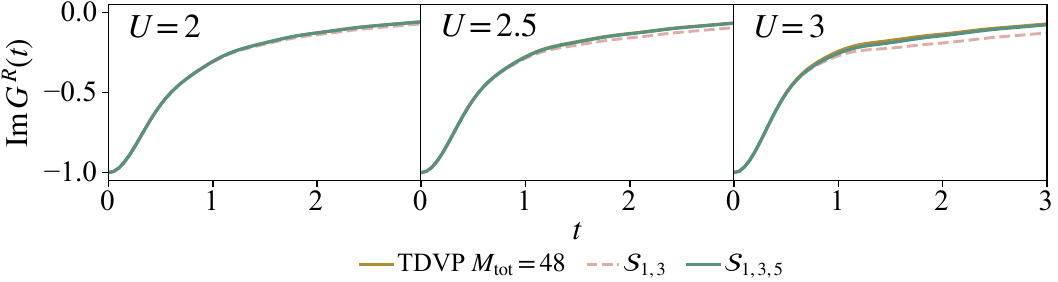}
\caption{The single-impurity $\operatorname{Im}G^R(t)$ at $\beta=\infty$ for $U=2,2.5,3$: $N_{\rm pm}=6$ coupled-Lindblad pseudomodes compared with the $M_{\rm tot}=48$ unitary-bath TDVP calculation.}
\label{fig:single-time}
\end{figure}

\emph{Single- and two-impurity Anderson models.}---The particle--hole-symmetric single-impurity Anderson model is given by
\begin{equation}
\hat  H_{\rm I} = -\epsilon \sum_{\sigma} \hat n_{\sigma} + U \hat n_{\uparrow}\hat n_{\downarrow},\quad \epsilon = U/2.
\end{equation}
The impurity couples to a semicircular bath with spectral density
$J (\omega)=\frac{1}{\pi}\sqrt{1-\frac{\omega^2}{W^2}}\mathbf{1}_{(-W,W)}(\omega)$ and half bandwidth $W=10$.
The lesser and greater hybridization functions are represented by $N_{\rm pm } = 2$ (\cref{fig:monomer-spectra}) and $N_{\rm pm } = 6$ (\cref{fig:single-time})
 coupled
pseudomodes per lesser or greater component. For $N_{\rm pm}=2$ and $\beta = 100$, we  compare C$^3$L with ED of the same full impurity-plus-pseudomode Lindbladian. These spectra are evaluated at $\eta=0$, with no added broadening. For $U=2$--$10$, the Krylov dimension $m_{\rm K}$ is capped at $128$ and $256$ in $\mc S_{1,3}$ and $\mc S_{1,3,5}$, respectively.
In the $U=10$, $\beta=1$ convergence test with $N_{\rm pm}=6$, the corresponding caps are $288$ and $384$.

\begin{figure}[!htbp]
\centering
\begin{subfigure}{\linewidth}
\centering
\includegraphics[width=0.65\linewidth]{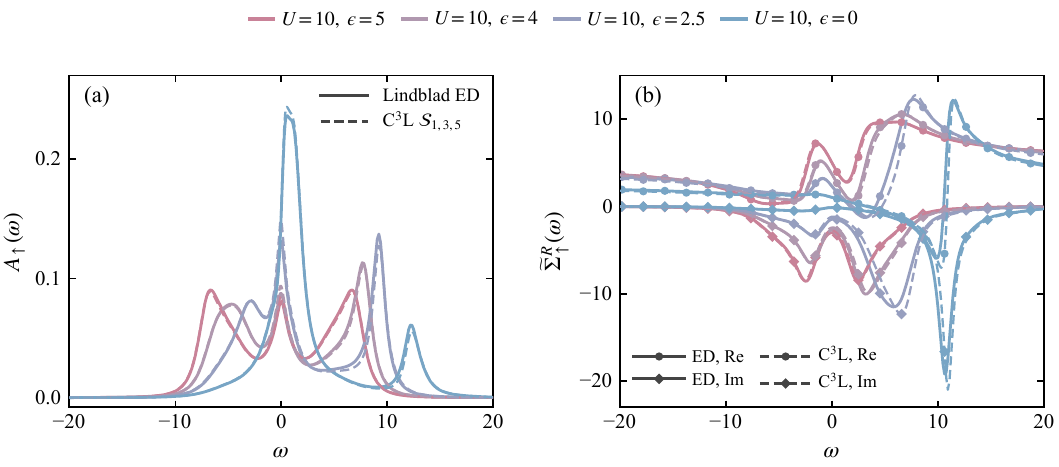}
\phantomsubcaption\label{fig:siam-filling-spectra}
\phantomsubcaption\label{fig:siam-filling-self-energy}
\end{subfigure}
\caption{Comparison of C$^3$L with $\mc S_{1,3,5}$ and Lindblad ED of the same fitted single-impurity model: (a) spectral functions $A_\uparrow(\omega)$; (b) real and imaginary parts of the full retarded self-energy $\wt\Sigma^R_\uparrow(\omega)$ (i.e. including the Hartree contribution), reconstructed through Dyson's equation using the same fitted bath. }
\label{fig:siam-filling-comparison}
\end{figure}

\begin{figure}[!htbp]
\centering
\includegraphics[width=0.75\linewidth]{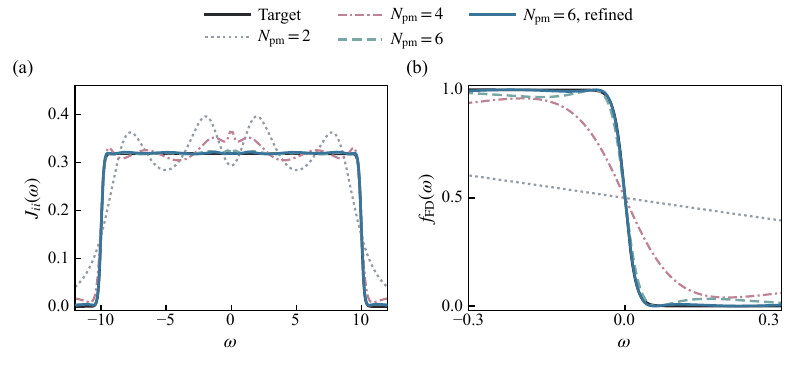}
\caption{Bath representations for the strongly correlated two-impurity benchmark at $\beta=100$, $\mu=0$, and $W=10$, with smooth cutoff width $0.1$. (a) Total bath spectral density. (b) Fermi--Dirac occupation $f_{\rm FD}(\omega)$ and its bath-fit approximations, obtained as the ratio of the occupied to total fitted spectral density $\frac{J_{\rm fit}^{\rm occ}(\omega)}{J_{\rm fit}^{\rm occ}(\omega)+ J_{\rm fit}^{\rm empty}(\omega)}$. The target (black) is compared with the initial $N_{\rm pm}=2,4,6$ fits and the refined $N_{\rm pm}=6$ fit used in Fig.~\ref{fig:kondo-benchmarks}. The frequency-focused refinement improves Fermi-edge resolution while increasing the time-window relative $L^2$ error from $1.05\%$ to $2.02\%$.}
\label{fig:two-impurity-bath-resolution}
\end{figure}

The $N_{\rm pm}=2$, $U=10$ benchmark at target $\beta=100$ tests the treatment of strong local Coulomb correlations. Although the qualitative strongly-correlated Hubbard-band structure remains visible, two pseudomodes per hybridization branch may not fully resolve the sharp Fermi edge and the associated long-time bath correlations, potentially weakening low-temperature Kondo features relative to the continuum impurity model~\cite{SchwarzGoldsteinDordaEtAl2016}. The benchmark with ED therefore assesses solver accuracy for this small pseudomode model, without establishing its low-energy continuum physics quantitatively. \cref{fig:siam-filling-comparison} compares the spectral functions and self-energies as the impurity level is shifted away from half filling. Across the fillings considered, the $\mc S_{1,3,5}$ response closely reproduces the ED spectral functions, while larger quantitative deviations appear in the self-energy away from half filling. The resolution of the Fermi edge is examined further in Fig.~\ref{fig:two-impurity-bath-resolution}, which compares different pseudomode numbers for the low-temperature two-impurity bath, including the refined $N_{\rm pm}=6$ representation.

Coupled-Lindblad pseudomodes can represent bath correlations over longer times with fewer modes~\cite{HuangParkChanLin2026,HuangDingWangEtAl2026}. To demonstrate this, we compare $N_{\rm pm}=6$ coupled pseudomodes with $M=48$ unitary-bath spin orbitals obtained by Gauss--Legendre discretization~\cite{DeVegaSchollwockWolf2015}. For a flat spectral density, this coincides with the Gauss-quadrature discretization underlying TEDOPA~\cite{PriorChinHuelgaEtAl2010}. For the semicircular baths used here, both give polynomially convergent unitary representations. The comparison is performed
at $\beta=\infty$, where the unitary-bath ground state $\ket{\psi_{\rm U}}$ can be computed efficiently with density matrix renormalization group (DMRG). The real-time Green's function can then be computed using TDVP with \texttt{ITensor}~\cite{FishmanWhiteStoudenmire2022}
\begin{equation}\label{eq:tdvp-retarded-green}
\begin{aligned}
[G^R_{\rm U}(t)]_{pq}=-\I\Theta(t)\big[
\langle\psi_{\rm U}|\hat a_p
e^{-\I t(\hat H_{\rm U}-E_{\rm U})}
\hat a_q^\dagger|\psi_{\rm U}\rangle +\langle\psi_{\rm U}|\hat a_q^\dagger
e^{\I t(\hat H_{\rm U}-E_{\rm U})}
\hat a_p|\psi_{\rm U}\rangle\big].
\end{aligned}
\end{equation}
Here $\hat H_{\rm U}$ is the full impurity--unitary-bath Hamiltonian and $E_{\rm U}$ its ground-state energy. The indices $p,q$ label impurity spin orbitals. TDVP evaluates the matrix elements by propagating the corresponding electron-addition and removal states.
For $U = 2,2.5$ and $3$, the retarded Green's function computed using C$^3$L at rank $\mathcal S_{1,3,5}$ agrees well with the TDVP unitary-bath result $G^R_{\rm U}(t)$.

\begin{figure}[!h]
\centering
\includegraphics[width=0.4\linewidth]{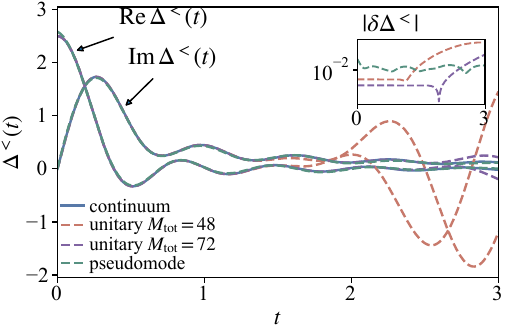}
\caption{Lesser hybridization function $\Delta^<(t)$ for the two-impurity benchmark at $\beta=\infty$: continuum bath, unitary baths with $M_{\rm tot}=48,72$ spin orbitals, and the coupled-Lindblad pseudomode fit with $N_{\rm pm}=4$. Both real and imaginary parts are shown. The inset gives the absolute complex error $\abs{\delta\Delta^<(t)} = |\Delta^<_{\rm fit}(t)-\Delta^<(t)|$ relative to the continuum bath.}
\label{fig:dimer-bath-fit}
\end{figure}

For the two-impurity Anderson model
\begin{equation}
\hat H_{\rm I} = - t_{12} \sum_{\sigma}(\hat a_{1\sigma}^\dag \hat a_{2\sigma} + \hat a_{2\sigma}^\dag \hat a_{1\sigma}) + U\sum_{i=1,2} \hat n_{i\uparrow}\hat n_{i\downarrow},
\end{equation}
we set $t_{12} = 0.5$, $U=2$ and chemical potential $\mu = U/2$. Each site couples to an independent bath at zero temperature $\beta  =\infty$, with the same semicircular spectral density $J(\omega)$ and half-bandwidth $W=10$ as in the single-impurity benchmark above. We use the coupled-Lindblad pseudomode representation with $N_{\rm pm} = 4$, corresponding to $2n = 2(n_{\rm I} + 2dN_{\rm pm}) = 36$. We independently construct a unitary bath representation with $M = 24$ and $36$ bath spin orbitals for each impurity site using the same discretization method, resulting in
$52$ and $76$ total spin orbitals for the unitary bath representation, respectively.
The corresponding hybridization fits are compared in Fig.~\ref{fig:dimer-bath-fit}. The $M_{\rm tot}=48$ unitary-bath fit loses accuracy around $t=1.5$, consistent with the loss of accuracy in the Green’s functions shown in \cref{fig:dimer}.

\emph{Imaginary-time reconstruction.}---At equilibrium, we evaluate the EOM-CC resolvent in \cref{eq:EOM-Green-function-frequency} at the Matsubara frequencies $\{\I\omega_n\}$ and reconstruct
\begin{equation}
G^{\rm M}_{pq}(\tau)=\frac{1}{\beta}\sum_{n\in\mathbb Z}
e^{-\I\omega_n\tau}\wt G^R_{pq}(\I\omega_n),
\quad \omega_n=\frac{(2n+1)\pi}{\beta}
\label{eq:matsubara-from-retarded}
\end{equation}
using the standard Matsubara summation; see Refs.~\cite[Appendix C]{BergeronHankevychKyungTremblay2011} and \cite[Appendix A]{DeeNakatsukasaWangJohnston2019}.

\emph{Continuum Dyson reconstruction.}---To remove the pseudomode-fitting error from the impurity response, we extract the C$^3$L self-energy with the fitted bath and re-embed it in the exact continuum hybridization $\Delta^R(z)$ of \cref{eq:bath_spectral_density_embedding}:
\begin{equation}\label{eq:continuum-dyson}
\wt\Sigma^R(z)=z-h-\Delta^R_{\rm fit}(z)-[\wt G^R(z)]^{-1},\quad
\wt G^R_{\rm cont}(z)=[z-h-\Delta^R(z)-\wt\Sigma^R(z)]^{-1},
\end{equation}
where $h$ is the impurity one-body matrix. In the examples considered, this reconstruction produces only small corrections to the directly computed impurity Green's functions.

\emph{Benchmark on the Kanamori model.}---For the interaction in \cref{eq:kanamori-hamiltonian}, we set $U=2$ and $J^{\rm H}=0.2$~\cite{WernerGullTroyerMills2008,YinHauleKotliar2011,DeMediciMravljeGeorges2011}. The chemical potential $\mu=(3U-5J^{\rm H})/2=2.5$ ensures half filling.
Following Refs.~\cite{EidelsteinGullCohen2020,KayeHuangStrandGolez2024,HuangKayeStrandGolez2025,YuErpenbeckZgidEtAl2025}, we use the metallic bath spectral density
\begin{equation}
  J_{ij}(\omega)=J^0_{ij}J_0(\omega)\quad \text{and} \quad J_0(\omega)=\frac{1}{\pi}\sqrt{1-\frac{\omega^2}{W^2}}\mathbf{1}_{(-W,W)}(\omega),
\end{equation}
where the half-bandwidth $W=2$, and the coupling $J^{0}_{ij} = \delta_{ij} + r(1-\delta_{ij})$ with $r=1$. We fit the lesser and greater hybridization components with $N_{\rm pm} = 5$ at both inverse temperatures $\beta = 8$ and $\beta = 64$.
For $r= 1$, the matrix $J^0 = \smqty(1&1\\1&1)$ has rank $1$, so only $d=1$ independent bath channel is required, giving
$2n = 2 (n_{\rm I}+ 2dN_{\rm pm})=  24$ spin orbitals.
The C$^3$L curves use the continuum Dyson reconstruction in \cref{eq:continuum-dyson}. For the spectra, we use the broadening $\omega\mapsto\omega+\I\eta$ with $\eta=0.05$. The Krylov dimension $m_{\rm K}$ is capped at $1024$ in both $\mc S_{1,3}$ and $\mc S_{1,3,5}$. On an NVIDIA H200 GPU, the combined runtime for computing the CCSD steady states, imaginary-time Green's functions, and real-frequency spectra is $114$~s ($115$~s) for $\mc S_{1,3}$ and $124$~s ($122$~s) for $\mc S_{1,3,5}$ at $\beta =8$ ($\beta = 64$).

\begin{figure}[!htbp]
\centering
\includegraphics[width=\linewidth]{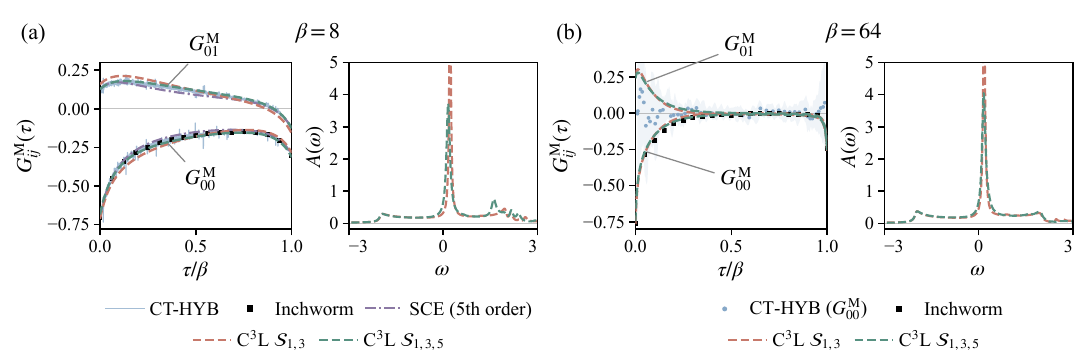}
{\phantomsubcaption\label{fig:kanamori-mu1-beta8}
\phantomsubcaption\label{fig:kanamori-mu1-beta64}}
\caption{Additional Kanamori results at $\mu=1$ (a) $\beta=8$; (b) $\beta=64$. For each temperature, the left plot shows diagonal and off-diagonal imaginary-time Green's functions, and the right plot shows the total spectral function per spin at $\eta=0.05$. Light shading in (b) reproduces the published CT-HYB uncertainty band from Ref.~\cite{EidelsteinGullCohen2020}.}
\label{fig:kanamori-mu1}
\end{figure}

\cref{fig:kanamori-mu1-beta8,fig:kanamori-mu1-beta64} show additional results away from half filling at $\beta=8,64$, respectively, both at $\mu=1$. We use the same $N_{\rm pm} = 5$ bath construction and spectral broadening $\eta=0.05$ as at half filling. The Krylov caps are $m_{\rm K} (\mc S_{1,3})= 1024$ and $m_{\rm K} (\mc S_{1,3,5}) = 1536$ at $\beta = 8$, and $m_{\rm K} (\mc S_{1,3})= 1024$ and $m_{\rm K} (\mc S_{1,3,5}) = 2048$ at $\beta = 64$. At $\beta=8$, the CT-HYB mean values for $G_{00}^{\rm M}$ and $G_{01}^{\rm M}$ are digitized from Ref.~\cite{KayeHuangStrandGolez2024}; the fifth-order strong coupling expansion (SCE) results of Ref.~\cite{HuangKayeStrandGolez2025} are taken from the numerical data accompanying Ref.~\cite{YuHarshaZhangEtAl2026}. The inchworm results at both temperatures and the CT-HYB results at $\beta=64$ are digitized from Ref.~\cite{EidelsteinGullCohen2020} and are for $G_{00}^{\rm M}$ only.

\emph{Benchmark against real-time inchworm QMC.}---For the general interacting Hamiltonian in \cref{eq:inchmoss-benchmark}~\cite{ErpenbeckBlommelZhangEtAl2024}, we use $\epsilon_0=V=V'=\nu'=-2 $, $U=\nu=2 $, and $\beta=0.1 $.
The environment consists of two identical equilibrium wide-band leads with combined spectral density
\begin{equation}
J_{ij}(\omega)=J_{ij}^{0}J_0(\omega)
\quad \text{and}\quad
J_0(\omega)=\frac{1}{\pi(1+e^{\upsilon(\omega-\omega_c)})(1+e^{-\upsilon(\omega+\omega_c)})},
\end{equation}
where $\upsilon  =10 $, $\omega_c=25 $, $J_{11}^{0}=J_{22}^{0}=3 /4$, and $J_{12}^{0}=J_{21}^{0}=1/4$. We use $N_{\rm pm} =5$ coupled-Lindblad pseudomodes for each lesser and greater hybridization component.
Since $J^0 = \smqty(3/4&1/4\\1/4&3/4)$ has rank 2, there are $d= 2$ independent bath channels, giving $2n  =2(n_{\rm I} + 2d N_{\rm pm}) = 44$ spin orbitals in total.
We evaluate the spectral function $A(\omega)$ at $\eta = 0$, with no added broadening. The Krylov dimension $m_{\rm K}$ is capped at $256$ in both $\mc S_{1,3}$ and $\mc S_{1,3,5}$.

\emph{Exchange-coupled molecular junction.}---For the Hamiltonian in \cref{eq:stm-exchange-hamiltonian} and the transport in \cref{fig:stm-exchange-transport}, we use four orbitals $(\mathrm{h}_2,\mathrm{h}_1,\mathrm{l}_1,\mathrm{l}_2)$ described by annihilation operators $\hat a_{i\sigma}$ at $\beta=10$. Here $\hat n_{i\sigma}=\hat a_{i\sigma}^\dagger\hat a_{i\sigma}$, $\hat n_i=\sum_\sigma\hat n_{i\sigma}$, and $\hat{\bv S}_i = (\hat S_{i}^x,\hat S_{i}^y,\hat S_{i}^z)$ with $ \hat S_i^\nu =\frac12\sum_{\sigma\sigma'}\hat a_{i\sigma}^\dagger\sigma^\nu_{ \sigma\sigma'}\hat a_{i\sigma'}$, where $ \sigma^\nu (\nu = x,y,z)$ are the Pauli matrices. The parameters are
\begin{equation}
h=\smqty(-2.95&0.05&0&0\\0.05&-2.55&0&0\\0&0&-0.71&0.04\\0&0&0.04&-0.48),\quad  U=0.8,\quad  V=0.3,\quad  J^{\rm E}=-J^{\rm P}=\begin{pmatrix}0.04&0.18\\0.25&0.04\end{pmatrix}.
\end{equation}
The rows of $J^{\rm E,P}$ are ordered as $(\mathrm{h}_2,\mathrm{h}_1)$ and the columns as $(\mathrm{l}_1,\mathrm{l}_2)$.
In the same original basis, the independent substrate ($\rm s$) and tip ($\rm t$) reservoirs have per-spin bath spectral-density matrices
\begin{equation}\label{eq:stm-contact-spectra}
J_r(\omega)=J_r^0 J_0(\omega-\mu_r)\bv u_r \bv u_r^\dagger ~( r \in \{\mathrm s,\mathrm t\}),\quad J_0(\omega)=\frac1\pi\sqrt{1-\frac{\omega^2}{W^2}}\,\mathbf{1}_{(-W,W)}(\omega),\quad W=4,\quad J_{\rm s}^0=0.10,\quad J_{\rm t}^0=0.01,
\end{equation}
with $\bv
u_{\rm s}=\frac{(1,1,1,1)^\top}2, \bv u_{\rm t}=\frac{(0.4,1,0.12,0.08)^\top}{\abs{(0.4,1,0.12,0.08)^\top}}, \mu_{\rm s}=0, \mu_{\rm t}=\Phi
$, so the tip preferentially extracts electrons from the HOMO-like orbital.
The total hybridization is $\Delta^{\gtrless}=\Delta^{\gtrless}_{\rm s}+\Delta^{\gtrless}_{\rm t}$, where
 $\Delta_r^{\gtrless}(t)=\int J_r(\omega)f_r^{\gtrless}(\omega)e^{-\I\omega t}\,\dInt\omega$, with $f_r^<=f_{\rm FD}^{\beta,\mu_r} $ and $f_r^>=1-f_{\rm FD}^{\beta,\mu_r} $. Bias shifts both the tip band and its occupation, with $\Phi<0$ driving extraction. For CC, $N_{{\rm pm},r}^<=N_{{\rm pm},r}^>=4$ for each reservoir, giving $2n = 40$ spin orbitals including the impurity. The tip matrices shift as $H_{\rm t}^{\gtrless}\mapsto H_{\rm t}^{\gtrless}+\Phi I$, while $\bv g_{\rm t}^{\gtrless}$ and $\Gamma_{\rm t}^{\gtrless}$ remain unchanged. The comparison methods use the continuum reservoirs. The energy-resolved current panels use $\Phi=-0.2,-0.4,\ldots,-1.6$ and the CC responses at $\eta=0$. The Krylov dimension $m_{\rm K}$ is capped at $1024$ and $1536$ in $\mc S_{1,3}$ and $\mc S_{1,3,5}$, respectively.

\emph{Molecular-orbital basis and equilibrium comparison.}---The fixed molecular operators $\hat f_{m\sigma}=\sum_i\Psi_{im}^*\hat a_{i\sigma}$ diagonalize the Fock matrix of the neutral isolated molecule. The index $m$ follows the order $(\mathrm{HOMO}-1,\mathrm{HOMO},\mathrm{LUMO},\mathrm{LUMO}+1)$.
In this basis, the Hamiltonian becomes
\begin{equation}
\hat H_{\rm I}=\sum_{mn\sigma}\wt h _{mn}\hat f_{m\sigma}^\dagger\hat f_{n\sigma}+\sum_{mnrs,\sigma\sigma'} \wt V  _{mnrs}\hat f_{m\sigma}^\dagger\hat f_{n\sigma'}^\dagger\hat f_{r\sigma'}\hat f_{s\sigma},
\end{equation}
where $\wt h =\Psi^\dagger h\Psi $ and $\wt V _{mnrs}=\sum_{ijkl}\Psi_{im}^*\Psi_{jn}^*V_{ijkl}\Psi_{kr}\Psi_{ls}$, with $V_{ijkl}$ the original-basis interaction coefficients in the main-text convention \cref{eq:impurity_hamiltonian}. %
The molecular-basis Green's functions are $G^{\rm M,MO}=\Psi^\dagger G^{\rm M}\Psi$. Thus the LUMO Green’s function is $G_{\rm LUMO}^{\rm M}=(G^{\rm M,MO})_{\rm LUMO,LUMO}$, associated with $\hat f_{\rm LUMO,\sigma}$. At equilibrium, it is reconstructed from 256 positive Matsubara frequencies using \cref{eq:matsubara-from-retarded}, with the leading $1/(\I\omega_n)$ tail included analytically (Refs.~\cite[Appendix C]{BergeronHankevychKyungTremblay2011} and \cite[Appendix A]{DeeNakatsukasaWangJohnston2019}).

  We diagonalize $\hat H_{\rm I}$ in the three-, four-, and five-electron sectors and resolve total spin within the eigenspaces, to obtain the many-body eigenvectors $\{\psi_i\}$. For the tip-contact operator $\hat b_{{\rm t}\sigma}=\sum_i (\mathbf u_{{\rm t}})_i\hat a_{i\sigma}$, the spin-summed removal spectrum from an initial multiplet $\mathcal M$ is
\begin{equation}\label{eq:stm-isolated-removal}
A^<_{{\rm iso},\mathcal M}(\omega)=\frac1{|\mathcal M|}\sum_{k\in\mathcal M,f,\sigma}|\langle \psi_f|\hat b_{{\rm t}\sigma}|\psi_k\rangle|^2\delta\!\left(\omega-(E_k-E_f)\right).
\end{equation}
Here $\mathcal M$ indexes the neutral singlet ground state or the equally weighted anionic ground-state doublet, and $f$ runs over eigenstates with one fewer electron.  The dashed sticks in \cref{fig:stm-mbpt-current,fig:stm-s13-current,fig:stm-s135-current} have heights given by these weights $|\langle \psi_f|\hat b_{{\rm t}\sigma}|\psi_k\rangle|^2$, summed over degenerate final states. For clarity, triplet and neutral-ionization lines below $10\%$ of the strongest line in their respective families are omitted.

\emph{Orbital-transfer strengths.}---In the original basis, define $\hat{\mathsf a}_i=(\hat a_{i\uparrow},\hat a_{i\downarrow})^\top$. The $\mathrm{l}_1\to \mathrm{h}_1$ orbital transfers decompose into a spin scalar and a spin vector~\cite{KanekoOhta2014,PokhilkoKrylov2019},
\begin{equation}\label{eq:stm-neutral-transfer}
\hat B_0=\frac1{\sqrt2}\hat{\mathsf a}_{\mathrm{h}_1}^\dagger I \hat{\mathsf a}_{\mathrm{l}_1},\quad \hat{  B}_1^\nu=\frac1{\sqrt2}\hat{\mathsf a}_{\mathrm{h}_1}^\dagger\sigma^\nu \hat{\mathsf a}_{\mathrm{l}_1}\quad (\nu = x,y, z).
\end{equation}
Their integrated steady-state strengths are
\begin{equation}\label{eq:stm-neutral-strengths}
W_0=\langle\hat B_0^\dagger\hat B_0\rangle=\frac 12 \langle \hat n_{\mathrm{l}_1}\rangle  -\frac14 \langle \hat n_{\mathrm{h}_1} \hat n_{\mathrm{l}_1}\rangle -\langle\hat{\bv S}_{\mathrm{h}_1}\cdot\hat{\bv S}_{\mathrm{l}_1}\rangle,\quad W_1=\sum_{\nu=x,y,z}\langle(\hat {B}_{1 }^\nu)^\dagger\hat B_{1 }^\nu \rangle=\frac{3}{2} \langle \hat n_{\mathrm{l}_1}\rangle -\frac34 \langle \hat n_{\mathrm{h}_1} \hat n_{\mathrm{l}_1}\rangle  + \langle \hat{\mathbf S}_{\mathrm{h}_1} \cdot\hat{\mathbf S}_{\mathrm{l}_1}\rangle .
\end{equation}
The ratio is defined as $R=\frac{W_1}{3W_0}$. Because $R$ depends on connected two-body correlations, it is not directly accessible from the one-particle Green's functions computed by $GW$ or 2ndB.
 In the sector with one electron on each orbital, the scalar and vector $\mathrm{l}_1\to \mathrm{h}_1$ transfers connect the singlet and triplet two-electron configurations, respectively, to the closed shell. %

For an observable $\hat O$ containing at most four fermionic operators, let $\bs O$ denote left multiplication by $\hat O$ and define $O_S=\bbra{I}\bs O\kett{\Phi_S}$, with $\kett{\Phi_0}=\kett{\rho_0}$. Its CCSD expectation value is
\begin{equation}\label{eq:stm-cc-observables}
\langle\hat O\rangle_{\rm CCSD}=\bbra{I}\bs O e^{\bs T}\kett{\rho_0}
=O_0+\sum_{|S|=2}O_S\,\mathrm t_S^{(2)}
+\sum_{|S|=4}O_S\eval{\left(\mathrm t^{(4)}+\frac12\mathrm t^{(2)}\wedge\mathrm t^{(2)}\right)}_S.
\end{equation}

\emph{Energy-resolved contact currents.}---All matrices below use the original basis. The per-spin removal and addition spectra are $A^<(\omega)=-\I \wt G^<(\omega)/(2\pi)$ and $A^>(\omega)=\I\wt G^>(\omega)/(2\pi)$. In terms of the injection and extraction kernels $\Xi_r^<$ and $\Xi_r^>$, the spin-summed injection current density is given by the Meir--Wingreen formula~\cite{MeirWingreen1992,JauhoWingreenMeir1994}:
\begin{equation}\label{eq:stm-current-density}
j_r(\omega)=2\Re\tr\!\left[\Xi_r^<(\omega)A^>(\omega)-\Xi_r^>(\omega)A^<(\omega)\right],\quad \mc J_r(\omega)=-j_r(\omega),\quad I_r=\int_\RR\mc J_r(\omega)\,\dInt\omega .
\end{equation}
Here $\mc J_r$ and $I_r$ are positive for extraction, as in the main text. For continuum reservoirs, $\Xi_r^{\gtrless}=2\pi f_r^{\gtrless}J_r$. For the fitted pseudomode reservoirs, we use the same $H_r^{\gtrless}$, $\Gamma_r^{\gtrless}$, and $\bv g_r^{\gtrless}$ as in \cref{eq:coupled-lindblad-hamiltonian,eq:dissipator}:
\begin{equation}\label{eq:stm-current-kernels}
\mathfrak g _r^{R,\gtrless}(\omega)=\left[\omega I-H_r^{\gtrless}+\I\Gamma_r^{\gtrless}\right]^{-1},\quad \mathfrak g _r^{A,\gtrless}=(\mathfrak g _r^{R,\gtrless})^\dagger,\quad \Xi_r^{\gtrless}=2(\bv   g _r^{\gtrless})^\dagger \mathfrak g _r^{R,\gtrless}\Gamma_r^{\gtrless}\mathfrak g_r^{A,\gtrless}\bv g_r^{\gtrless}.
\end{equation}
The static extraction current is evaluated independently from the full spin-orbital 1-RDM $\gamma_{pq}=\langle\hat c_q^\dagger\hat c_p\rangle$ (see \cref{sec:HF}). Let ${\rm I}$ denote the impurity spin orbitals and $\mc I_r$ the pseudomode spin orbitals of reservoir $r$, including both branches and spins. Then
\begin{equation}\label{eq:stm-static-current}
I_r=-2\Im\sum_{p\in{\rm I},\,q\in\mc I_r}\mathsf H_{pq}\gamma_{qp}.
\end{equation}
Here $\mathsf H$ is the one-body coefficient matrix of $\hat H_{\rm CL}$.

The combined runtime for computing the CCSD steady states, the imaginary-time Green's functions and real-frequency spectra on an NVIDIA H200 GPU is $122$~s for $\mc S_{1,3}$ and $25$~min for $\mc S_{1,3,5}$.

\emph{Strongly correlated two-impurity Anderson model with exchange.}---For the benchmark in Fig.~\ref{fig:kondo-benchmarks}, we consider $J^{\rm E}=2$ and $0$, $U=4$ and $\beta=100$. Each impurity couples to an independent bath with half-bandwidth $W=10$ and a smooth cutoff of width $0.1$,
\begin{equation}
J_{ij}(\omega)=\frac{\delta_{ij}}{\pi[1+e^{(\omega-W)/0.1}][1+e^{(-\omega-W)/0.1}]}\approx \frac{\delta_{ij}}\pi \Theta(W-\abs{\omega}).
\end{equation}
For each impurity-spin channel, we use $N_{\rm pm}=6$ coupled-Lindblad pseudomodes per lesser or greater branch. The full two-impurity pseudomode model contains $52$ spin orbitals: $4$ impurity and $48$ pseudomode spin orbitals. We first fit the bath on $t\in[0,100]$ with \texttt{RealTimeBath}, then refine the bath matrices separately using \texttt{SciPy}'s trust-region reflective nonlinear least-squares solver~\cite{VirtanenGommersOliphantEtAl2020,BranchColemanLi1999}, with \texttt{max\_nfev=3000} in the final stage. For one channel, we write $J=J_{ii}$ and define the fitted frequency-domain functions
\begin{equation}\label{eq:bath-refinement-resolvents}
\Delta_{\rm fit}^{\gtrless}(z)=(\bv g^{\gtrless})^\dagger(zI-H^{\gtrless}+\I\Gamma^{\gtrless})^{-1}\bv g^{\gtrless},\quad \Delta_{\rm fit}^R=\Delta_{\rm fit}^<+\Delta_{\rm fit}^>,\quad J_{\rm fit}^{\rm occ}(\omega)=-\frac{1}{\pi}\Im\Delta_{\rm fit}^<(\omega),\quad J_{\rm fit}^{\rm empty}(\omega)=-\frac{1}{\pi}\Im\Delta_{\rm fit}^>(\omega).
\end{equation}
The corresponding time-domain hybridization is $\Delta_{\rm fit}^<(t)=(\bv g^<)^\dagger e^{(-\I H^<-\Gamma^<)t}\bv g^<$. At each refinement stage, we minimize
\begin{equation}\label{eq:bath-frequency-refinement}
\begin{aligned}
\min_{\theta} \frac{1}{|\Omega|}\sum_{\omega\in\Omega}w_\omega\left[\left|\Delta_{\rm fit}^R(\omega)-\Delta^R(\omega)\right|^2+\pi^2\left|J_{\rm fit}^{\rm occ}(\omega)-J(\omega)f_{\rm FD}^{100,0}(\omega)\right|^2\right]
 +\chi_{\rm t}\frac{\sum_{t\in\mc T}w_t\left|\Delta_{\rm fit}^<(t)-\Delta^<(t)\right|^2}{\sum_{t\in\mc T}w_t\left|\Delta^<(t)\right|^2}+\chi_{\rm r}\norm{\theta-\theta_0}_2^2.
\end{aligned}
\end{equation}
Here, the exact retarded hybridization is the Cauchy transform of the bath spectral density. The inverse relation follows from the Plemelj formula:
\begin{equation}\label{eq:bath_spectral_density_embedding}
 \Delta^R(z)=\int_{-\infty}^{\infty}\frac{J(\epsilon)}{z-\epsilon}\dd\epsilon\quad (\Im z>0),\quad
 J(\omega)=\frac{\I}{2\pi}\lim_{\eta \downarrow 0 }\left[\Delta^R(\omega+\I\eta)-\Delta^R(\omega+\I\eta )^\dagger\right].
\end{equation}
We set $\chi_{\rm t}=0.04$ and $\chi_{\rm r}=10^{-12}$ for the time-domain penalty and parameter regularization, respectively. The empirical $\ell^2$ weights $w_\omega=1+15e^{-(\omega/0.3)^2}$ and $w_t=1+t/2$ emphasize the Fermi edge and long times. The real parameter vector $\theta$ comprises the independent real components of Hermitian $H^<$ and the real and imaginary parts of $\Upsilon$ and $\bv g^<$, with $\Gamma^<=\Upsilon\Upsilon^\dagger\succeq0$; $\theta_0$ denotes the starting vector for each refinement stage. Particle--hole symmetry fixes $H^>=-(H^<)^*$, $\Gamma^>=(\Gamma^<)^*$, and $\bv g^>=(\bv g^<)^*$. The frequency grid $\Omega$ combines uniform grids of $481$, $161$, and $161$ points on $[-12,12]$, $[-2,2]$, and $[-0.2,0.2]$, respectively. The time grid is $\mc T=\{0,0.025,\ldots,2\}\cup\{0,0.2,\ldots,200\}$.

Spectra are evaluated in the $\mc S_{1,3,5}$ EOM space with broadening $\eta = 0.01$ and Krylov dimension $m_{\rm K} = 768$. For $\abs{\omega}\le 0.5$, we instead use restarted GMRES with $m_{\rm K}=256$ and at most $8$ cycles for $0.1<\abs{\omega}\le 0.5$, and $m_{\rm K}=768$ and at most $8$ cycles for $\abs{\omega}\le 0.1$.
\cref{fig:two-impurity-bath-resolution} compares the total bath spectral density and Fermi-edge resolution for different numbers of pseudomodes.

\section{A simple mechanism for MBPT descriptions of exchange-induced multiplet Green's functions}\label{app:multiplet_GF_mbpt}

We present an isolated two-orbital model with closed-form exact, self-consistent 2ndB and $GW$ Green's functions. We will observe that the missing physical interaction in those MBPT approximations can lead to qualitatively incorrect descriptions of the exchange-induced multiplet structure.

The Hamiltonian and reference state are
\begin{equation}\label{eq:mbpt-toy-H}
  \hat H=\epsilon\hat n_1
  +U(\hat n_{1\uparrow}\hat n_{1\downarrow}+\hat n_{2\uparrow}\hat n_{2\downarrow})
  -J^{\rm E}\left(2\hat{\mathbf S}_1\cdot\hat{\mathbf S}_2+\frac12\hat n_1\hat n_2\right),
  \quad \ket{\psi}=\ket{2\uparrow},\quad \epsilon,U,J^{\rm E}>0.
\end{equation}
The one-particle reference $\ket{\psi}$ is an eigenstate with energy zero. It may also be interpreted as a minimal one-hole representation of a three-electron molecular HOMO--LUMO system, with particle addition representing electron removal.
The Hamiltonian conserves $\hat n_1,\hat n_2$ and $\hat S^z$. Since the reference has definite orbital occupations and spin, the Green's function is diagonal:
 $
  G^{\gtrless}_{i\sigma,j\sigma'}(t)
  =\delta_{ij}\delta_{\sigma\sigma'}G^{\gtrless}_{i\sigma}(t),
   i,j=1,2. $
Using the definitions
$G^>_p(t)=-\I\mel{\psi}{\hat a_p(t)\hat a_p^\dagger}{\psi}$,
$G^<_p(t)=\I\mel{\psi}{\hat a_p^\dagger\hat a_p(t)}{\psi}$, and
$G^R_p(t)=\Theta(t)[G^>_p(t)-G^<_p(t)]$, we can first evaluate the addition and removal processes that reach a single eigenstate. The vacuum, $\ket{2\uparrow 2\downarrow}$, and $\ket{1\uparrow,2\uparrow}$ are eigenstates with energies $0$, $U$, and $\epsilon-J^{\rm E}$, respectively. Their time evolution contributes a single phase factor, giving
\begin{equation}\label{eq:table_green}\text{\small
\begin{tabular}{ccccc}
\toprule
Spin orbital & State reached by the nonzero branch & $G^>(t)$ & $G^<(t)$ & $\wt G^R(z) (\Im z>0)$\\
\midrule
$2\uparrow$ & Removal to $\ket{0}$ & $0$ & $\I$ & $\frac{1}{z}$  \\
$2\downarrow$ & Addition to $\ket{2\uparrow 2\downarrow}$ & $-\I e^{-\I Ut}$ & $0$ & $\frac{1}{z-U}$ \\
$1\uparrow$ & Addition to $\ket{1\uparrow,2\uparrow}$ & $-\I e^{-\I(\epsilon-J^{\rm E})t}$ & $0$ &$ \frac{1}{z-(\epsilon-J^{\rm E})}$  \\
\bottomrule
\end{tabular}}
\end{equation}
For the remaining $1\downarrow$ entry, adding $1\downarrow$ to $\ket{\psi}=\ket{2\uparrow}$ reaches $\ket{\mathrm{p}}=\ket{1\downarrow,2\uparrow}$. This is not an eigenstate because exchange couples it to $\ket{\mathrm{q}}=\ket{1\uparrow,2\downarrow}$. That is, $ \ket{\psi}\xrightarrow{\hat a^\dagger_{1\downarrow}}\ket{\mathrm{p}}\xleftrightarrow{-J^{\rm E}}\ket{\mathrm{q}}$. The $\ket{\rm p}$, $\ket{\rm q}$ configurations nevertheless form a closed subspace.
The symmetric triplet $\ket{\bv T_0}=(\ket{\mathrm{p}}+\ket{\mathrm{q}})/\sqrt{2}$ and antisymmetric singlet $\ket{\bv S}=(\ket{\mathrm{p}}-\ket{\mathrm{q}})/\sqrt{2}$ are eigenstates with energies $\epsilon-J^{\rm E}$ and $\epsilon+J^{\rm E}$, respectively. Note that $\hat a_{1\downarrow}\ket{\psi}=0$, $\hat H\ket{\psi}=0$, and $\hat a^\dagger_{1\downarrow}\ket{\psi}=\ket{\mathrm{p}}$. Thus, only the addition branch contributes to $G^R_{1\downarrow}$.
The invariance of the $\{\ket{\mathrm{p}},\ket{\mathrm{q}}\}$ subspace and $\ket{\mathrm{p}}=(\ket{\bv T_0}+\ket{\bv S})/\sqrt{2}$ give
\begin{equation}
  G^>_{1\downarrow}(t)
  =-\frac{\I}{2}\left[e^{-\I(\epsilon-J^{\rm E})t}+e^{-\I(\epsilon+J^{\rm E})t}\right],
  \quad G^<_{1\downarrow}(t)=0.
\end{equation}
The Fourier transform of $G^R_{1\downarrow}(t)$ then gives the exact frequency-domain Green's function
\begin{equation}\label{eq:mbpt-toy-exact}
  \wt G^R_{1\downarrow}(z)
  =\frac{1}{z-\epsilon-(J^{\rm E})^2/(z-\epsilon)}
  =\frac12\left[\frac{1}{z-\epsilon+J^{\rm E}}+\frac{1}{z-\epsilon-J^{\rm E}}\right].
\end{equation}
The $1\downarrow$ spectrum has two equally weighted peaks separated by $2J^{\rm E}$, independently of $U$. We next examine why 2ndB and $GW$ miss this structure.

We use the interaction convention
\begin{equation}\label{eq:mbpt-toy-tensor}
  \hat H_{\mathrm{int}}
  =\frac12\sum_{ijkl,\sigma\tau}\mathsf{C}_{ij,kl}
  \hat a^\dagger_{i\sigma}\hat a^\dagger_{k\tau}\hat a_{l\tau}\hat a_{j\sigma},
  \quad \mathsf{C}_{11,11}=\mathsf{C}_{22,22}=U,\quad \mathsf{C}_{12,21}=\mathsf{C}_{21,12}=J^{\rm E},
\end{equation}
with all other entries zero. The spin-orbital interaction elements are
$V_{(i\sigma)(k\tau),(j\sigma')(l\tau')}=\mathsf{C}_{ij,kl}\delta_{\sigma\sigma'}\delta_{\tau\tau'}$, and the antisymmetrized interaction is
$\overline V_{\alpha\beta,\gamma\delta}=V_{\alpha\beta,\gamma\delta}-V_{\alpha\beta,\delta\gamma}$.
We separate the HF contribution from the retarded self-energy as $\wt\Sigma^R(z)=\Sigma^{\mathrm{HF}}+\wt\Sigma^c(z)$, where
$\Sigma^{\mathrm{HF}}_{\alpha\beta}=\sum_\gamma\overline V_{\alpha\gamma,\beta\gamma}\mel{\psi}{\hat n_\gamma}{\psi}=\overline V_{\alpha\,2\uparrow,\beta\,2\uparrow}$.
In spin-orbital order $(1\uparrow,1\downarrow,2\uparrow,2\downarrow)$, this gives $\Sigma^{\mathrm{HF}}=\operatorname{diag}(-J^{\rm E},0,0,U)$. Adding the bare one-particle energies $\epsilon_{1\sigma}=\epsilon$ and $\epsilon_{2\sigma}=0$ gives the HF energies $f_p=\epsilon_p+\Sigma^{\mathrm{HF}}_{pp}$, namely
$f_{1\uparrow}=\epsilon-J^{\rm E}$, $f_{1\downarrow}=\epsilon$, $f_{2\uparrow}=0$, and $f_{2\downarrow}=U$.
For each diagonal element, Dyson's equation reconstructs the Green's function as $\wt G^R_p(z)=[z-f_p-\wt\Sigma^c_p(z)]^{-1}$. In particular,
 $
  \wt G^R_{1\downarrow}(z)
  =\frac{1}{z-\epsilon-\wt\Sigma^c_{1\downarrow}(z)}.
 $
Here $\wt\Sigma^c(z)$ is the retarded correlation contribution, which we approximate using 2ndB or $GW$. \cref{eq:mbpt-toy-exact} gives the exact correlation self-energy,
\begin{equation}\label{eq:exact_selfenergy}
  \wt\Sigma_{1\downarrow}^c(z)=\frac{(J^{\rm E})^2}{z-\epsilon}.
\end{equation}

\emph{Second Born.}---For diagonal propagators $G^{\gtrless}$, the complete self-consistent 2ndB correlation term is~\cite{StefanucciVanLeeuwen2013,RomanielloBechstedtReining2012}
\begin{equation}\label{eq:mbpt-toy-2B-full}
  \Sigma^{c,\gtrless}_{\alpha\beta}(t)=\Sigma^{(2),\gtrless}_{\alpha\beta}(t)
  =\frac12\sum_{\gamma\mu\nu}
  \overline V_{\alpha\gamma,\mu\nu}\overline V_{\mu\nu,\beta\gamma}
  G^{\gtrless}_\mu(t)G^{\gtrless}_\nu(t)G^{\lessgtr}_\gamma(-t).
\end{equation}
For the greater component, $G^<_\gamma$ is nonzero only for $\gamma=2\uparrow$, as shown in \cref{eq:table_green}. With $\gamma=2\uparrow$, the only contributing pairs $(\mu,\nu)$ are $(1\uparrow,2\downarrow)$ and $(2\downarrow,1\uparrow)$, with interaction vertices $-J^{\rm E}$ and $+J^{\rm E}$, respectively. This gives
\begin{equation}\label{eq:mbpt-toy-2B}
  \Sigma^{(2),>}_{1\downarrow}(t)
  =(J^{\rm E})^2G^>_{1\uparrow}(t)G^>_{2\downarrow}(t)G^<_{2\uparrow}(-t)
  =-\I (J^{\rm E})^2e^{-\I(\epsilon+U-J^{\rm E})t}.
\end{equation}
Every other greater entry vanishes by the same interaction-vertex selection. All lesser entries require $(\mu,\nu)=(2\uparrow,2\uparrow)$ and therefore vanish by antisymmetry of $\overline V$. The Fourier transform of the retarded kernel $\Sigma^c(t)=\Theta(t)[\Sigma^{c,>}(t)-\Sigma^{c,<}(t)]=\Theta(t)\Sigma^{(2),>}(t)$ yields
\begin{equation}\label{eq:mbpt-toy-2B-frequency}
  \wt\Sigma^c_{1\downarrow,\text{2ndB}}(z)
  =\frac{(J^{\rm E})^2}{z-(\epsilon+U-J^{\rm E})}.
\end{equation}
Substituting this closed-form self-energy into Dyson's equation gives the 2ndB Green's function,
\begin{equation}\label{eq:mbpt-toy-2B-G}
  \wt G^R_{1\downarrow,\text{2ndB}}(z)
  =\frac{1}{z-\epsilon-(J^{\rm E})^2/[z-(\epsilon+U-J^{\rm E})]}.
\end{equation}
Both its poles and residues depend on $U$, in contrast to the exact response in \cref{eq:mbpt-toy-exact}.

\begin{figure}[htbp]
  \centering
  \includegraphics[width=0.75\linewidth]{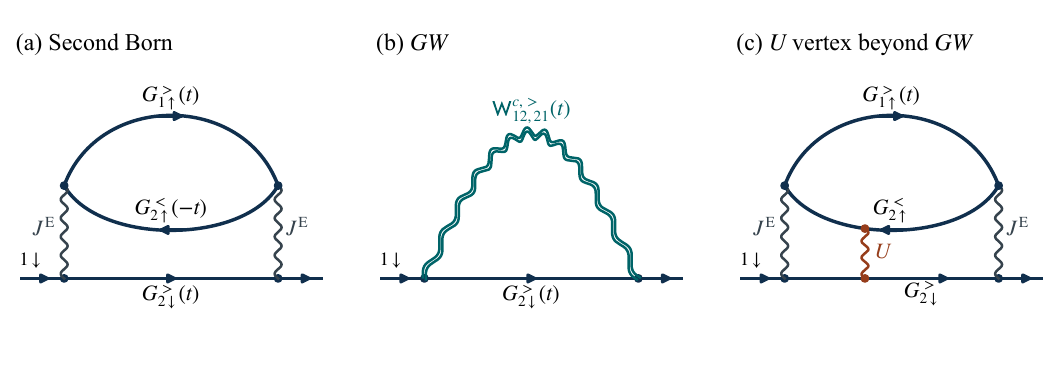}
  \caption{Self-energy diagrams for $1\downarrow$ addition. Solid arrows denote bold propagators, a single wavy line denotes a bare interaction, and the double wavy line denotes the correlation part of the screened interaction, $\mathsf{W}^c=\mathsf{W}-\mathsf{C}$. The labels identify the contributing spin-orbital channels. (a) The surviving direct 2ndB diagram\footnote{The crossed diagram vanishes because it requires equal spins on the external and occupied internal lines, here $1\downarrow$ and $2\uparrow$. } propagates $1\uparrow$ and $2\downarrow$ particles and a $2\uparrow$ hole between two exchange vertices, omitting both intermediate particle--hole energy corrections, $+J^{\rm E}$ and $-U$. (b) $GW$ resums the $1\uparrow$ particle--$2\uparrow$ hole pair inside $\mathsf{W}^c$, supplying the $+J^{\rm E}$ correction, while the $2\downarrow$ particle still propagates separately. (c) A representative $U$ insertion, omitted by both $GW$ and 2ndB, couples the $2\downarrow$ particle to the $2\uparrow$ hole.}
  \label{fig:mbpt-multiplet-diagrams}
\end{figure}

The physical picture of this failure is clear. The intermediate configuration is
$\ket{\mathrm{q}}=\hat a^\dagger_{1\uparrow}\hat a^\dagger_{2\downarrow}\hat a_{2\uparrow}\ket{\psi}$, which contains two particle lines and one hole line, denoted $2p1h$ (see \cref{eq:mbpt-toy-2B} and Fig.~\ref{fig:mbpt-multiplet-diagrams}(a)). However, the two $J^{\rm E}$ vertices in 2ndB are already needed to create and remove $\ket{\mathrm{q}}$ from $\ket{\rm p}$. Between them, the three lines can only propagate independently, without additional vertices connecting them. 2ndB therefore assigns the energy $f_{1\uparrow}+f_{2\downarrow}-f_{2\uparrow}=\epsilon+U-J^{\rm E}$, the self-energy pole in \cref{eq:mbpt-toy-2B-frequency}. Adding $2\downarrow$ independently costs $U$ because $2\uparrow$ is occupied, while adding $1\uparrow$ independently gains an exchange $-J^{\rm E}$. In $\ket{\mathrm{q}}$, however, that $2\uparrow$ particle has been removed. The spurious charging cost $U$ and exchange gain $-J^{\rm E}$ therefore arise because 2ndB omits interactions among the intermediate lines.

\emph{$GW$.}---The screening approximation uses~\cite{StefanucciVanLeeuwen2013,RomanielloBechstedtReining2012}
\begin{equation}\label{eq:gw_c}
  \Sigma^{c,\gtrless}_{ij,\sigma}(t)
  =\I\sum_{kl}G^{\gtrless}_{kl,\sigma}(t)\mathsf{W}^{c,\gtrless}_{ik,lj}(t).
\end{equation}
Here $\mathsf W$ denotes the screened interaction, $\mathsf W^c$ its correlation part, and $\mathsf P$ the polarization, defined by
\begin{equation}\label{eq:mbpt-toy-GW-def}
  \mathsf{P}^{\gtrless}_{ij,kl}(t)
  =-\I\sum_\sigma G^{\gtrless}_{jk,\sigma}(t)G^{\lessgtr}_{li,\sigma}(-t),
  \quad \mathsf{W}=\mathsf{C}+\mathsf{C}\ast\mathsf{P}\ast\mathsf{W},\quad
  \mathsf{W}^c=\mathsf{W}-\mathsf{C}.
\end{equation}
Here $\ast$ denotes convolution on the Keldysh--Schwinger contour, with $\mathsf C$ acting as an instantaneous kernel~\cite{StefanucciVanLeeuwen2013}. The screening equation becomes $\wt{\mathsf W}^R=\mathsf C+\mathsf C\wt{\mathsf P}^R\wt{\mathsf W}^R$ in the frequency domain with
\begin{equation}
  (\mathsf{C}\wt{\mathsf{P}}^R\wt{\mathsf{W}}^R)_{ij,kl}(\omega)
  =\sum_{ab,cd}\mathsf{C}_{ij,ab}\wt{\mathsf{P}}^R_{ab,cd}(\omega)\wt{\mathsf{W}}^R_{cd,kl}(\omega).
\end{equation}
There are only two nonzero polarization entries, associated with $2\uparrow\to1\uparrow$ according to \cref{eq:table_green,eq:mbpt-toy-GW-def}
\begin{equation}\label{eq:mbpt-toy-P}
  \wt{\mathsf{P}}^R_{21,12}(\omega)=\frac{1}{\omega-(\epsilon-J^{\rm E})},\quad
  \wt{\mathsf{P}}^R_{12,21}(\omega)=-\frac{1}{\omega+(\epsilon-J^{\rm E})},\quad \Im\omega>0.
\end{equation}
Thus, for $\mathsf C=\mathsf C_U+\mathsf C_{J^{\rm E}}$, one has $\mathsf C_U\mathsf P=\mathsf P\mathsf C_U=0$ by \cref{eq:mbpt-toy-tensor}. The two surviving screening equations are
\begin{equation}
  \wt{\mathsf{W}}^R_{12,21}(\omega)=J^{\rm E}+\frac{J^{\rm E}}{\omega-(\epsilon-J^{\rm E})}\wt{\mathsf{W}}^R_{12,21}(\omega),\quad
  \wt{\mathsf{W}}^R_{21,12}(\omega)=J^{\rm E}-\frac{J^{\rm E}}{\omega+(\epsilon-J^{\rm E})}\wt{\mathsf{W}}^R_{21,12}(\omega).
\end{equation}
Subtracting the bare interaction $\mathsf C$ gives
\begin{equation}\label{eq:mbpt-toy-W}
  \wt{\mathsf{W}}^{c,R}_{12,21}(\omega)=\frac{(J^{\rm E})^2}{\omega-\epsilon},\quad
  \wt{\mathsf{W}}^{c,R}_{21,12}(\omega)=-\frac{(J^{\rm E})^2}{\omega+\epsilon},\quad
  \mathsf{W}^{c,>}_{12,21}(t)=-\I (J^{\rm E})^2e^{-\I\epsilon t}.
\end{equation}
The only nonzero contribution to \cref{eq:gw_c} closes $\mathsf W^c$ on the remaining $2\downarrow$ particle, as shown in Fig.~\ref{fig:mbpt-multiplet-diagrams}(b).
\begin{equation}\label{eq:mbpt-toy-GW}
\begin{aligned}
  \Sigma^{c,>}_{1\downarrow}(t)
=\I G^>_{2\downarrow}(t)\mathsf{W}^{c,>}_{12,21}(t)
   =-\I (J^{\rm E})^2e^{-\I(\epsilon+U)t},\quad
  \Sigma^{c,<}_{1\downarrow}(t)=0,\quad
  \wt\Sigma^c_{1\downarrow,GW}(z)=\frac{(J^{\rm E})^2}{z-(\epsilon+U)}.
\end{aligned}
\end{equation}
Similar to \cref{eq:mbpt-toy-2B-G}, $GW$ places the intermediate pole at $\epsilon+U$ instead of $\epsilon$, making the Green's-function poles and residues depend spuriously on $U$.
Physically, screening correctly assigns energy $\epsilon$ to the neutral excitation from $\ket{2\uparrow}$ to $\ket{1\uparrow}$. However, $GW$ still then combines it with an independently added $2\downarrow$ particle of energy $U$. It removes the spurious $J^{\rm E}$ gain but keeps the $U$ cost associated with the $2\uparrow$ particle that should be absent in $\ket{\mathrm{q}}$. Recovering the exact self-energy \cref{eq:exact_selfenergy} requires interactions between the $2\downarrow$ particle and the $2\uparrow$ hole, as illustrated in Fig.~\ref{fig:mbpt-multiplet-diagrams}(c).

\begin{figure}[t]
  \centering
  \includegraphics[width=0.45\linewidth]{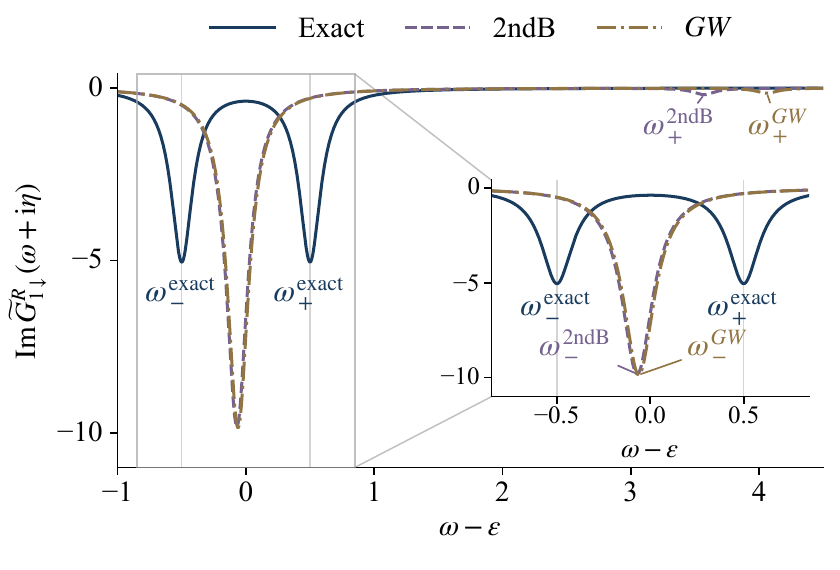}
  \caption{$\Im \wt G^R_{1\downarrow}(\omega+\I\eta)$ for the exact, self-consistent 2ndB, and self-consistent $GW$ responses at $\epsilon=6$, $U=4$, $J^{\rm E}=0.5$, and $\eta=0.1$. The full frequency range shows the dominant approximate resonances and their very weak satellites. The inset shows the singlet--triplet region, with equally weighted exact resonances at $\omega-\epsilon=\pm J^{\rm E}$ and the dominant 2ndB/$GW$ resonance.}
  \label{fig:mbpt-multiplet-spectra}
\end{figure}

\emph{Consequences for the multiplet spectral features.}---We discuss the consequence of the missing qualitative interactions on the spectral features. Since
the other three propagators retain their HF poles and updating $G_{1\downarrow}$ leaves the nonzero self-energy contractions unchanged, both solutions are already self-consistent, with
\begin{equation}\label{eq:mbpt-toy-auxiliary}
  \wt G^R_{1\downarrow}(z)
  =\frac{1}{z-\epsilon-(J^{\rm E})^2/(z-\epsilon-a)}
  =\left(\left[zI-\begin{pmatrix}\epsilon&-J^{\rm E}\\-J^{\rm E}&\epsilon+a\end{pmatrix}\right]^{-1}\right)_{11},
\end{equation}
where $a=0$ for the exact solution, $a=U-J^{\rm E}$ for 2ndB, and $a=U$ for $GW$. The approximate self-energies shift the intermediate pole by an amount of order $U$. The poles and residues of $\wt G^R_{1\downarrow}$ are
\begin{equation}\label{eq:mbpt-toy-poles}
  \omega_\pm=\epsilon+\frac{a\pm\sqrt{a^2+4(J^{\rm E})^2}}{2},\quad
  \mathrm{Res}_{\omega_\pm }\wt G^R_{1\downarrow}=\frac12\left(1\mp\frac{a}{\sqrt{a^2+4(J^{\rm E})^2}}\right).
\end{equation}
\cref{fig:mbpt-multiplet-spectra} illustrates the resulting responses at $\epsilon=6$, $U=4$, and $J^{\rm E}=0.5$. Both approximations produce two poles, but replace the equally weighted exchange peaks with a dominant resonance and an extremely weak, misplaced satellite.

Recovering the exact self-energy pole requires correlations among the different pairs of the $2p1h$ configuration. Correlating the $1\uparrow$ particle--$2\uparrow$ hole and $2\downarrow$ particle--$2\uparrow$ hole pairings within the same $2p1h$ propagator supplies the $+J^{\rm E}$ and $-U$ corrections that restore the intermediate energy to $\epsilon$~\cite{RomanielloBechstedtReining2012,DegrooteVanNeckBarbieri2011,LischnerDeslippeJainLouie2012}.
Hamiltonian EOM-CCSD accommodates those interactions~\cite{ShavitBartlett2009} by explicitly including both $\ket{\mathrm{p}}$ and $\ket{\mathrm{q}}$ in its $1p/2p1h$ response space. For the molecular junction, this example illustrates a possible missing correlation that can arise even without a nonequilibrium bath.

\makeatletter
\let\@FMN@list\@empty
\patchcmd{\NAT@bibsetnum}{LastBibItem}{LastSIBibItem}{}{}
\patchcmd{\endthebibliography}{LastBibItem}{LastSIBibItem}{}{}
\makeatother
\end{bibunit}
\end{document}